\documentclass[prd,preprintnumbers,amsmath,amssymb,floatfix]{revtex4}
\usepackage{amsmath,amsfonts,latexsym,amssymb,graphicx,graphics,epsfig,subfigure,color,makeidx,cases}
\usepackage{xcolor,diagbox,enumitem}
\usepackage{multirow}
\usepackage[colorlinks,linkcolor=blue,anchorcolor=blue,citecolor=green,urlcolor=blue]{hyperref}
\usepackage{mathrsfs}
\renewcommand{\baselinestretch}{1.25} 

\begin{document}
\title{Decoding Brane Splitting: Constraints from Quasinormal Spectra}

\author{Hai-Long Jia$^{a}$$^{b}$\footnote{jiahl2024@lzu.edu.cn}}
\author{Yu-Xiao Liu$^{a}$$^{b}$\footnote{liuyx@lzu.edu.cn, corresponding author}}

\affiliation{
$^{a}$Lanzhou Center for Theoretical Physics, 
    Key Laboratory of Theoretical Physics of Gansu Province, 
    Key Laboratory of Quantum Theory and Applications of the Ministry of Education, 
    Gansu Provincial Research Center for Basic Disciplines of Quantum Physics,
    Lanzhou University, Lanzhou 730000, China \\
$^{b}$Institute of Theoretical Physics \& Research Center of Gravitation,
    School of Physical Science and Technology, 
    Lanzhou University, Lanzhou 730000, China 
}

\begin{abstract}

The internal structure of a thick braneworld can generate a double-barrier
effective potential for tensor perturbations, allowing the quasinormal mode
(QNM) spectrum to encode information about brane splitting.
However, extracting barrier scales from finite spectral data requires
clarifying the model-dependent relation between spectral features and effective-potential geometry.
Using a symmetric double P\"{o}schl--Teller potential as a reference,
we quantify how finite sets of QNM frequencies constrain the parameters
of split-braneworld effective potentials and assess the associated systematic errors.
We derive a closed-form expression for the quadratic frequency shift
at small barrier separation and fixed overtone number, use a high-frequency
relation within the independent-barrier approximation to obtain initial parameter estimates, 
and test selected QNM sequences of the full potential by direct integration
with outgoing boundary conditions.
For a family of analytic braneworld background solutions, we distinguish
among geometric barrier parameters (peak height, width, and separation),
reference scales constructed from the positions and strengths of complex poles of the potential, 
and effective potential parameters obtained by fitting selected modes.
In the two split-brane spectra examined at higher overtones,
the ray estimates of the inverse width and barrier height move toward
the complex-pole benchmarks as the overtone window shifts upward.
For the $s=19$ Bernstein input, an equal-weight fit predicts the complex
frequencies of four modes excluded from the fit with a relative
root-mean-square (RMS) error below $1\%$.
The relative error in the fundamental-mode damping rate is
approximately $35\%$.
These results quantify the accuracy of inferring split-brane effective-potential scales from 
finite QNM data and identify model-dependent limits.

\end{abstract}

\maketitle
\tableofcontents 
\section{Introduction} \label{Introduction}

Braneworld gravity involves both the recovery of four-dimensional gravity on the brane 
and the dynamics of massive gravitational excitations.
Models with large or warped extra dimensions provide a geometric framework
for studying the hierarchy problem and the scale dependence of gravity
\cite{ArkaniHamed:1998millimeter,RandallSundrum:1999hierarchy,RandallSundrum:1999alternative,MaartensKoyama:2010}.
In thick braneworlds, the brane is described as a domain wall of finite
thickness supported by smooth bulk matter fields.
Its internal structure and dynamics can be characterized through background
solutions and perturbation analysis
\cite{DeWolfe:1999,Gremm:2000gravity,Gremm:2000thick,Dzhunushaliev:2010thick,Liu:2017extra}.
The form of the perturbation equations depends on the gravitational theory
and the perturbation sector, as illustrated by studies of tensor and scalar
perturbations in warped $f(R)$ models~\cite{ZhongLiu:2017fR}.
Here we focus on tensor gravitational perturbations, whose equations reduce,
after separation of variables, to a one-dimensional Schr\"{o}dinger-type spectral problem.
The normalizable graviton zero mode corresponds to four-dimensional gravity
localized on the brane, while massive Kaluza--Klein (KK) modes represent
gravitational excitations along the extra dimension
\cite{Csaki:2000universal,ClarksonSeahra:2005}.
Imposing outgoing boundary conditions on the massive sector determines
the quasinormal modes (QNMs) and their complex frequencies, whose real
and imaginary parts characterize oscillation and decay, respectively
\cite{Seahra:2005ringing,Seahra:2005metastable,Berti:2009qnm}.
The QNM spectrum thus links the dynamics of these excitations to the
extra-dimensional effective potential and provides a probe of the brane's
internal structure.

In thick-brane backgrounds with a split structure, the tensor effective
potential can take a double-barrier form.
Multiple reflections and partial trapping between the barriers can produce
long-lived resonances, beating, and delayed signals.
Previous studies of thick and double braneworlds have mainly investigated
these effects by computing spectra and waveforms for prescribed backgrounds
\cite{Tan:2022sound,Tan:2024characteristic,Tan:2024dbl,Zhu:2025echoes,Deng:2025scalar}.
Resonance studies in braneworlds also include the ringdown of black holes
localized on the brane~\cite{Chung:2016qnm}, 
whereas we focus on KK resonances governed by the effective potential along the extra dimension.
However, long-lived resonances alone do not establish the formation of
a double-barrier cavity; the spectral structure and trapping conditions
of the specific background must also be considered.
Related studies have examined quasi-localization in Rastall gravity,
a continuum mass gap in de Sitter branes, 
and threshold trapping in models with a finite extra dimension, 
illustrating the spectral properties of gravitational excitations in different backgrounds
\cite{Tan:2024Rastall,Jia:2025dS,Jia:2024finite}.
Studies of graviscalar modes and their late-time tails also show that
the spectrum and time-domain response depend on the perturbation sector \cite{Tan:2025graviscalar}.
Moreover, deformations of the central well can alter wave trapping even
when the barrier separation changes little.
These considerations motivate a further question in the massive tensor sector: 
which effective-potential scales can a finite set of QNM frequencies
constrain, and to what extent do these constraints reflect the brane's split structure?

Answering this question requires choosing a reference family of potentials
and specifying its range of applicability.
Previous studies of symmetric double barriers and wormholes have developed
methods for extracting effective geometric parameters from QNM spectra
\cite{Voelkel:2018inverse,VoelkelKokkotas:2018wormhole,Konoplya:2018wormhole,Voelkel:2020thesis},
but a finite set of QNM frequencies is insufficient to uniquely determine
an arbitrary effective potential \cite{Zworski:2017resonances}.
Furthermore, the spectral problem with outgoing boundary conditions
is non-self-adjoint: small potential perturbations can produce large shifts
in individual QNM frequencies without inducing comparable changes in the dominant time-domain signal
\cite{Jaramillo:2021pseudospectrum,Jaramillo:2022gwsignature,Berti:2022stability,Cheung:2022flea,Ianniccari:2024deciphering}.
Such spectral sensitivity has been demonstrated for quantum-corrected
Schwarzschild and Boulware--Deser--Wheeler black holes \cite{Cao:2024pseudospectrum,Cao:2025BDW};
studies of semi-open systems show that changes in boundary reflectivity
can affect the QNM branch structure \cite{Wu:2026semiopen}.
Studies of thick braneworlds have likewise found that sensitivity
of the QNM spectrum to perturbations can coexist with relatively robust
early ringdown \cite{Jia:2026butterfly}.

In this work, we use the established analytic thick-brane backgrounds of
Ref.~\cite{Tan:2024dbl} to develop and test a systematic procedure
for assessing the accuracy of effective-potential inference from
finite QNM spectra and diagnosing the sources of error.
The symmetric reference potential combines two identical
P\"{o}schl--Teller (PT) barriers. The component height, width, and
center-to-center separation can be varied independently, and the
isolated-barrier scattering amplitudes are known analytically
\cite{PoschlTeller:1933,Ferrari:1984pt,Bueno:2018echoes}.
We derive a closed-form quadratic frequency shift at small
separation and fixed overtone number, and use a high-frequency
scattering relation for independent barriers to initialize the
inversion. Direct integration by shooting, with outgoing boundary
conditions, tests these approximations for the full summed potential.
Synthetic spectra with known parameters calibrate the finite-window
bias of the high-frequency inversion. Shifting the mode window tests
parameter stability, and the local sensitivity of the frequencies
indicates how well each parameter is constrained. Varying the input
frequencies and fitting weights tests sensitivity to the numerical
inputs and to the fitting objective. We check predictions for QNMs
excluded from the fit and assess the fundamental-mode damping rate
with its own error measure. Compactly supported deformations that
preserve the algebraic tail probe the sensitivity of this damping
rate to the outer potential profile. Moving to higher overtones can
reduce the finite-window bias of the ray inversion and increase the
sensitivity of some inferred parameters to errors in the input spectrum.
We compare the inferred parameters with two benchmarks derived
directly from the effective potential. The real-axis peak positions,
heights, and curvatures determine the geometric benchmark. The
positions and strengths of the nearest complex poles determine the
second benchmark. Tracking absolute fractional deviations from both
benchmarks across equal-size overtone windows shows which physical
scales the estimates approach. In the two split-brane spectra tested
at higher overtones, the ray estimates of inverse width and height
move toward the complex-pole scales and away from the real-axis
peak scales.

Section~\ref{sec:effective-potentials} introduces the tensor effective
potential and spectral features of brane splitting;
Sec.~\ref{sec:dpt-spectrum} studies the analytic spectral structure
of the double-PT potential;
Sec.~\ref{sec:spectral-inversion} presents the finite-spectrum constraint
method and its validation;
Sec.~\ref{sec:applications} analyzes spectral constraints and systematic
errors in braneworld backgrounds; and
Sec.~\ref{conclusion} gives the conclusions.
Numerical methods and reproducibility details are collected in the Appendix.

\section{Effective Potentials and Spectral Features of Brane Splitting} \label{sec:effective-potentials}

We consider transverse-traceless (TT) tensor perturbations $h_{\mu\nu}^{\rm TT}$ of a
five-dimensional thick-brane background, 
following the notation of Ref.~\cite{Jia:2024finite}.
In the conformal coordinate $z$, the background metric with flat
four-dimensional slices takes the form
\begin{equation}
    ds^2=e^{2A(z)}
    \left(\eta_{\mu\nu}dx^\mu dx^\nu+dz^2\right),
    \label{metric-conformal}
\end{equation}
Here $\eta_{\mu\nu}=\operatorname{diag}(-1,1,1,1)$ is the
Minkowski metric, and the warp function $A(z)$ gives the warp factor
$e^{2A(z)}$.
After Fourier decomposition along the three spatial directions on the brane,
the canonically normalized tensor master amplitude $\Psi_{\rm gt}(t,z)$ satisfies
\begin{equation}
    \label{wave-equation-vgt}
    \left[-\partial_t^2+\partial_z^2-V_{\rm gt}(z)-p^2\right]
    \Psi_{\rm gt}(t,z)=0,
\end{equation}
where $p^2=\delta^{ij}p_i p_j$ is the squared magnitude of the spatial momentum on the brane.
For the Einstein backgrounds considered below, 
this master variable is defined through the decomposition
$h_{\mu\nu}^{\rm TT} =e^{-3A/2}\epsilon_{\mu\nu}e^{ip_i x^i}\Psi_{\rm gt}(t,z)$,
where $\epsilon_{\mu\nu}$ is a polarization tensor satisfying the TT conditions.
Setting $\Psi_{\rm gt}(t,z)=e^{-i\omega t}\psi_{\rm gt}(z)$,
Eq.~\eqref{wave-equation-vgt} reduces to
\begin{equation}
    \left[-\partial_z^2+V_{\rm gt}(z)\right]
    \psi_{\rm gt}(z)=m^2\psi_{\rm gt}(z),
    \qquad m^2=\omega^2-p^2.
    \label{schrodinger-vgt}
\end{equation}
For KK normal modes with real mass, 
$m$ denotes the four-dimensional KK mass; 
in the resonance problem, it is a complex spectral parameter determined by outgoing boundary conditions \cite{Berti:2009qnm}.
Throughout the following, we set $p=0$ and adopt the convention $m=\omega$.

The tensor spectral operator considered here admits a supersymmetric factorization \cite{Cooper:1994eh}.
Introducing a real superpotential $\mathcal{W}(z)$ determined by the background, we define
\begin{equation}
    \label{susy-operators}
    \mathcal{D}=\partial_z+\mathcal{W},
    \qquad
    \mathcal{D}^{\dagger}=-\partial_z+\mathcal{W}.
\end{equation}
Here $\dagger$ denotes the formal adjoint with respect to the measure $dz$.
The original tensor operator and its supersymmetric partner are
\begin{equation}
    \label{susy-factorization}
    \mathcal{D}\mathcal{D}^{\dagger}
    =-\partial_z^2+V_{\rm gt},
    \qquad
    \mathcal{D}^{\dagger}\mathcal{D}
    =-\partial_z^2+V_{\rm gt}^{\rm dual},
\end{equation}
where
\begin{equation}
    V_{\rm gt}=\mathcal{W}^2+\partial_z\mathcal{W},
    \qquad
    V_{\rm gt}^{\rm dual}=\mathcal{W}^2-\partial_z\mathcal{W}.
    \label{dual-potentials}
\end{equation}
For backgrounds in five-dimensional Einstein gravity with minimally 
coupled matter fields, $\mathcal{W}=3\partial_zA/2$, and Eq.~\eqref{dual-potentials} gives
\begin{equation}
    V_{\rm gt}=\frac{3}{2}\partial_z^2A
    +\frac{9}{4}(\partial_zA)^2,
    \qquad
    V_{\rm gt}^{\rm dual}=-\frac{3}{2}\partial_z^2A
    +\frac{9}{4}(\partial_zA)^2.
    \label{vgt-definition}
\end{equation}
We refer to $V_{\rm gt}^{\rm dual}$ as the ``dual potential'' throughout.
It is the supersymmetric partner potential associated with the same background.
Its spectral correspondence with the original tensor potential
is determined jointly by the intertwining map and the boundary conditions.

Denote the wave function for the dual potential by $\bar{\psi}_{\rm gt}$.
For $m\neq0$, solutions of the original and dual problems are related by
\begin{equation}
    \label{susy-intertwining}
    \bar{\psi}_{\rm gt}
    =\frac{1}{m}\mathcal{D}^{\dagger}\psi_{\rm gt},
    \qquad
    \psi_{\rm gt}
    =\frac{1}{m}\mathcal{D}\bar{\psi}_{\rm gt}.
\end{equation}
We use the analytic family of thick-brane backgrounds in Ref.~\cite{Tan:2024dbl}, 
where $s$ is a positive odd integer, $k>0$ is an inverse-length scale, 
and $\delta$ is a dimensionless thickness parameter.
We set $\delta=1$, for which the warp function is
\[
    A_s(z)=-\frac{1}{2s} \ln\left[1+(kz)^{2s}\right].
\]
As $|z|\to\infty$, this background satisfies
$\mathcal{W}(z)\sim-3/(2z)\to0$.
Consequently, $m^{-1}\mathcal{D}^{\dagger}e^{\pm imz}\sim\mp i e^{\pm imz}$,
so the intertwining map changes only the asymptotic amplitudes
of outgoing waves while preserving their outgoing character.
With the same analytic continuation prescription for outgoing solutions,
the original and dual potentials therefore have identical nonzero
KK quasinormal frequencies \cite{Cooper:1994eh,Jia:2024finite}.

The localized graviton zero mode must be treated separately because
Eq.~\eqref{susy-intertwining} does not apply at $m=0$.
The first-order equations $\mathcal{D}^{\dagger}\psi_0=0$
and $\mathcal{D}\bar{\psi}_0=0$ yield the formal zero-energy solutions
$\psi_0\propto\exp(\int\mathcal{W}dz)$ and $\bar{\psi}_0\propto\exp(-\int\mathcal{W}dz)$, respectively.
In Einstein gravity, these become $e^{3A/2}$ and $e^{-3A/2}$.
For the backgrounds considered here, only the former is normalizable
according to $\int|\psi|^2dz<\infty$.
The dual potential thus retains the nonzero resonance poles of interest
while excluding the normalizable graviton zero mode of the original
potential.

The QNM spectrum of the dual potential is determined by purely outgoing
conditions at both spatial infinities:
\begin{equation}
    \label{qnm-boundary-condition}
    \bar{\psi}_{\rm gt}(z)\sim
    \begin{cases}
        C_+e^{+imz}, & z\rightarrow+\infty,\\
        C_-e^{-imz}, & z\rightarrow-\infty,
    \end{cases}
\end{equation}
where $C_\pm$ are asymptotic amplitudes.
For complex frequencies, these conditions are defined by 
analytic continuation of the outgoing scattering solutions.
The dual potential of this background has the algebraic tail
$V_{\rm gt}^{\rm dual}\sim3/(4z^2)$,
so Eq.~\eqref{qnm-boundary-condition} specifies the leading asymptotic
terms of the outgoing solutions.
The numerical calculations use outgoing conditions consistent with this
tail, as detailed in the Appendix.
With the time dependence $e^{-imt}$, decaying QNMs satisfy
$\operatorname{Im}m<0$, and their amplitude damping rate is
$\gamma=-\operatorname{Im}m$.
For nonzero spatial momentum on the brane, the corresponding complex
frequency can be recovered from $\omega^2=m^2+p^2$,
choosing the branch continuously connected to $\omega=m$ at $p=0$.

\begin{figure*}[htb]
\centering
\includegraphics[width=0.98\textwidth]{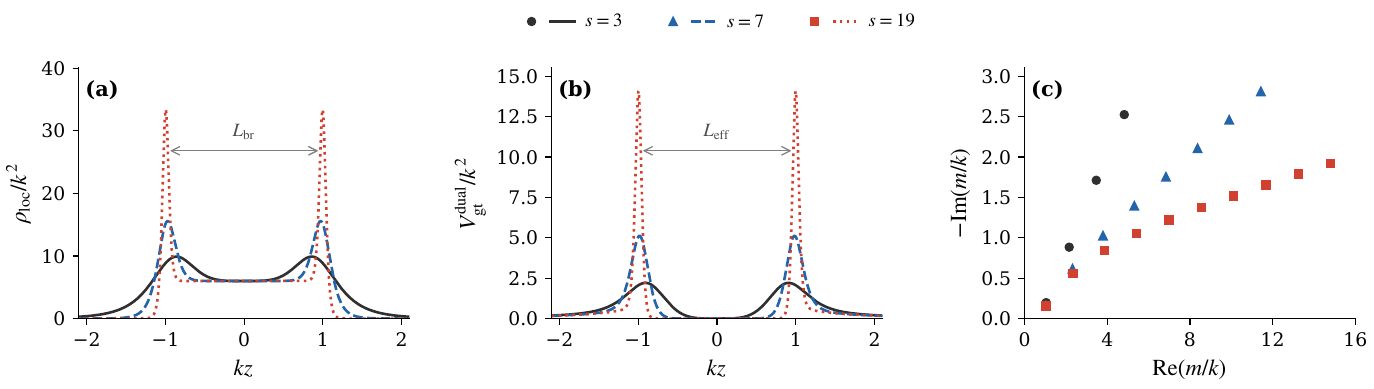}
\caption{
Brane splitting and numerical tensor spectra for the analytic backgrounds
of Ref.~\cite{Tan:2024dbl}, with $s=3,7,19$ and $k=\delta=1$.
(a) Energy density with the asymptotic background value subtracted.
(b) Tensor dual potentials.
(c) Numerically computed quasinormal frequencies on the branch with
positive real part:
$n=0,\ldots,3$ for $s=3$,
$n=0,\ldots,7$ for $s=7$,
and $n=0,\ldots,9$ for $s=19$.
All markers show QNM frequencies obtained by direct integration.
$L_{\rm br}$ and $L_{\rm eff}$ indicate the separations of the
energy-density peaks and the dual-potential peaks, respectively,
for $s=19$.
Here $L_{\rm eff}$ corresponds to $L_{\rm geo}$ defined in the text.
For $s=3$, the displayed frequencies change by less than
$5\times10^{-11}k$ when the integration configuration is varied.
}
\label{fig:brane-splitting-dual-potential}
\end{figure*}

We now examine the effective potential structure reflected in these resonance frequencies.
For this background family, the dual potential in the unsplit case $s=1$ has a single main barrier.
For $s>1$, it develops two main peaks, with a region between them
that can support partial wave trapping.
For the background family considered here, brane splitting is manifested
in both the energy-density profile and the double-barrier structure
of the dual tensor potential, as shown in
Fig.~\ref{fig:brane-splitting-dual-potential}.
To characterize the effective scattering geometry, we extract local
double-barrier parameters directly from the dual potential and use them
as benchmarks independent of spectral fitting.
Let $z_-$ and $z_+$ denote the locations of the two dominant,
nondegenerate maxima with positive heights, and define
\begin{equation}
    \label{barrier-descriptors}
    V_\pm=V_{\rm gt}^{\rm dual}(z_\pm),
    \qquad
    \alpha_\pm^2
    =-\frac{\partial_z^2V_{\rm gt}^{\rm dual}(z_\pm)}
    {2V_\pm},
    \qquad
    L_{\rm geo}=z_+-z_-.
\end{equation}
Here $V_\pm$ are the peak heights of the full potential,
$\alpha_\pm^{-1}$ are local width scales defined by the peak curvatures,
and $L_{\rm geo}$ is the peak separation.
This convention ensures that the local PT reference potential
$V_\pm\operatorname{sech}^2[\alpha_\pm(z-z_\pm)]$
matches the height and second derivative of the original potential at each peak.
For reflection-symmetric backgrounds, the two peaks have equal heights
and local widths, and the corresponding parameters are denoted by
$V_{\rm pk}$, $\alpha_{\rm loc}$, and $L_{\rm geo}$.
These quantities are determined directly from the known potential;
their relation to parameters estimated from finite spectral data
must be tested through independent comparison.

Since the tensor wave equation is written in the conformal coordinate $z$,
$L_{\rm geo}$ provides an optical length scale set by the geometric
separation of the two reflection regions.
The actual resonance phase and echo delay also depend on the potential
profile and the reflection phase.
Moreover, the relation between $L_{\rm geo}$ and the proper separation
of sub-branes or the parameters of the matter-field configuration
depends on the specific background solution.
We therefore treat $L_{\rm geo}$ as a geometric descriptor of the dual
potential and discuss its connection to brane splitting within
a specified model.

Throughout, $n=0$ labels the fundamental mode of each spectrum
and $n=1,2,\ldots$ label its successive overtones.
For spectra tabulated with one-based mode numbers, we subtract one
from the published index.

Figure~\ref{fig:brane-splitting-dual-potential} compares the background
structure and tensor spectra for $s=3,7,19$ and $k=\delta=1$.
Figure~\ref{fig:brane-splitting-dual-potential}(a) shows the energy density
after subtraction of its asymptotic background value,
$\rho_{\rm loc}=\rho-\rho_\infty$.
We follow the normalization $\kappa_5^2=1$ of Ref.~\cite{Tan:2024dbl},
for which $\rho_\infty=-6k^2$.
Figure~\ref{fig:brane-splitting-dual-potential}(b) shows the dual potentials,
and Fig.~\ref{fig:brane-splitting-dual-potential}(c) shows the
numerically computed quasinormal frequencies within the plotted range
for the corresponding potentials.

For the backgrounds shown, the real frequency increases along each
displayed QNM sequence.
The real frequencies of the fundamental modes are relatively close
across the different backgrounds, whereas the higher modes exhibit
more pronounced differences in both oscillation frequency and damping rate.
The distribution of several overtones therefore provides information
about the potential profile beyond that contained in a single
fundamental-mode frequency.
By contrast, in the unsplit-brane example of Ref.~\cite{Tan:2024dbl},
the real parts of the first ten QNM frequencies decrease successively.

These spectral features can also be understood in terms of double-barrier
scattering.
Waves propagate between the barriers and undergo repeated partial
reflections.
The propagation and reflection phases jointly determine the resonance
condition, while transmission through the barriers allows waves to leak out.
This structure can support long-lived KK resonances and produce delayed
echoes under suitable excitation.
When several resonances with nearby frequencies are excited simultaneously,
their interference can also produce beating
\cite{Tan:2024dbl,Deng:2025scalar}.
These phenomena illustrate different time- and frequency-domain features
of double-barrier scattering.

To analyze this scattering problem quantitatively, we adopt the symmetric
double P\"{o}schl--Teller (DPT) potential as a three-parameter reference model:
\begin{equation}
    V_{\rm DPT}(z)
    =V_b\,\operatorname{sech}^2\!\left[
    \alpha_{\rm PT}\left(z-\frac{L}{2}\right)\right]
    +V_b\,\operatorname{sech}^2\!\left[
    \alpha_{\rm PT}\left(z+\frac{L}{2}\right)\right],
    \label{double-pt-potential}
\end{equation}
where
$V_b$ is the height of each component barrier,
$\alpha_{\rm PT}^{-1}$ is its width scale,
and $L$ is the separation between the component barrier centers.
We also write the full potential as $V_L(z)\equiv V_{\rm DPT}(z)$.
These component parameters must be distinguished from the peak parameters
of the full potential.
When the barriers overlap significantly, the peak heights, local widths
defined by the peak curvatures, and peak separation of the full potential
generally differ from the corresponding component parameters.
In particular, the full potential may have only a single peak.
In this case, the double-peak separation $L_{\rm geo}$ is no longer
applicable, whereas the component center separation $L$ remains
a well-defined parameter of the reference model.

The double-PT reference model and implicit Gamma-function resonance
conditions constructed from analytic single-barrier scattering amplitudes
have been used in earlier studies of the inverse QNM problem for symmetric
double barriers \cite{VoelkelKokkotas:2018wormhole,Voelkel:2020thesis,Bueno:2018echoes}.
Within this established framework, we distinguish the independent-barrier
approximation from the spectral problem for the full summed potential,
analyze low-overtone behavior in different separation regimes,
derive spectral relations within a high-frequency scattering model,
and assess their range of applicability using numerical spectra
of the full potential and applications to braneworlds.

\begin{figure*}[htb]
    \centering
    \includegraphics[width=0.95\textwidth]{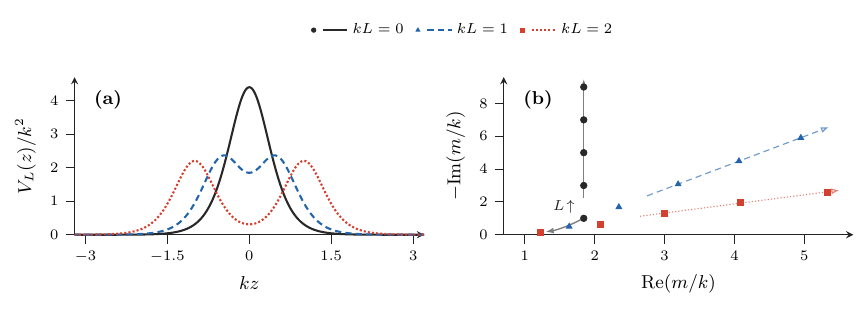}
    \caption{
    Representative spectral features of the symmetric double-PT potential
    in Eq.~\eqref{double-pt-potential}, with
    $V_b=2.2k^2$, $\alpha_{\rm PT}=2k$, and $kL=0,1,2$.
    (a) Potential profiles.
    At $kL=2$, the peak height of the full potential is close to that
    of the $s=3$ dual potential in
    Fig.~\ref{fig:brane-splitting-dual-potential}(b),
    and their peak separations are comparable,
    although the PT component barriers are wider.
    (b) Five low-lying quasinormal frequencies computed using the
    Bernstein polynomial spectral method \cite{Fortuna:2023bernstein},
    on the branch with positive real part.
    The gray curved arrow indicates the direction of motion
    of the fundamental mode as $L$ increases.
    The straight arrows in matching colors indicate increasing overtone
    number.
    For $kL=1,2$, the lines are least-squares fits to the three highest
    overtones shown and illustrate the trends within these finite
    overtone windows.
    }
    \label{fig:double-barrier-spectral-features}
\end{figure*}

Figure~\ref{fig:double-barrier-spectral-features} illustrates the relevant
spectral features of the symmetric double-PT potential with
$V_b=2.2k^2$ and $\alpha_{\rm PT}=2k$.
At $kL=2$, the peak height of the full potential is close to that of
the $s=3$ dual potential in
Fig.~\ref{fig:brane-splitting-dual-potential}(b),
and their peak separations are comparable,
although the chosen PT component barriers are wider.
This comparison illustrates qualitative features of double-barrier
scattering and does not constitute a precise fit to the braneworld potential.
At $L=0$, the two terms combine into a single PT barrier of height $2V_b$.
For the chosen parameters, the quasinormal frequencies on the branch
with positive real part share the same real part, forming a vertical
sequence of poles.
Over the displayed range of separations, the gray curved arrow in
Fig.~\ref{fig:double-barrier-spectral-features}(b) schematically indicates
the motion of the fundamental mode toward a lower real frequency
and a smaller damping rate as $L$ increases, reflecting the formation
of longer-lived cavity resonances.
The higher QNM frequencies shown form approximately linear sequences
in the complex-frequency plane, with orientations that depend on $L$.
The straight arrows in matching colors indicate the trends of these
finite sequences.
For $kL=1,2$, the lines are obtained by linear fits to the three highest
overtones shown.
The corresponding high-frequency scattering relation will be derived
in Sec.~\ref{subsec:high-overtone}.
The quasinormal frequencies in
Fig.~\ref{fig:double-barrier-spectral-features}(b)
are computed using the Bernstein polynomial spectral method
\cite{Fortuna:2023bernstein}.

The double-PT reference potential uses three parameters to describe selected
features of braneworld spectra.
Within this reference family, the inverse problem aims to constrain
effective-potential parameters for a specified set of QNMs and accuracy
targets, predict quasinormal frequencies excluded from the fit,
and compare the resulting spectral descriptors with benchmarks extracted
independently from the effective potential.
This procedure requires separate assessments of numerical errors
in the forward spectral calculations and model errors introduced
by the reference-potential approximation.
The next section presents the small-separation expansion at fixed overtone
number and analytic relations within the high-frequency scattering model.
Section~\ref{sec:spectral-inversion} then explains how these relations
are used to construct initial parameter estimates for spectral fitting.
The reliability of the inferred constraints is assessed by computing
frequencies for the full potential and testing predictions on QNMs
excluded from the fit.

\section{Analytic Spectral Structure of the Double P\"{o}schl--Teller Potential}
\label{sec:dpt-spectrum}

To discuss its spectral structure at different separations,
we first specify the relevant length scale.
A single component barrier,
$V_b\operatorname{sech}^{2}[\alpha_{\rm PT}(z-z_0)]$,
has the characteristic width
\begin{equation}
    \label{eq:dpt-intrinsic-width}
    w_{\rm PT}=\alpha_{\rm PT}^{-1}.
\end{equation}
For example, its full width at half maximum is
$2\operatorname{arccosh}\sqrt{2}\,w_{\rm PT}$;
other commonly used definitions of width likewise differ from
$w_{\rm PT}$ only by a constant factor.
The dimensionless separation
$\ell\equiv L/w_{\rm PT}=\alpha_{\rm PT}L$ characterizes the degree
of barrier overlap. In braneworld units,
$\ell=(\alpha_{\rm PT}/k)(kL)$, so
$\ell=2(kL)$ for the parameters in
Fig.~\ref{fig:double-barrier-spectral-features}.
The midpoint value of the full potential,
\begin{equation}
    \label{eq:dpt-midpoint-overlap}
    V_L(0)
    =2V_b\operatorname{sech}^{2}
    \left(\frac{\ell}{2}\right)
    \sim8V_b e^{-\ell}
    \qquad(\ell\gg1),
\end{equation}
measures the suppression of overlap at large separation.

Our analysis at different separations is based on the same
transfer-matrix resonance condition
\cite{Mostafazadeh:2020transfer,Zworski:2017resonances}.
For real $m>0$, we write
\begin{equation}
    \label{eq:dpt-transfer-matrix}
    \psi(z)
    =A_{\pm}e^{imz}+B_{\pm}e^{-imz}+o(1),
    \qquad z\to\pm\infty, \qquad
    \binom{A_+}{B_+}
    =M(m;V_b,\alpha_{\rm PT},L)\binom{A_-}{B_-}.
\end{equation}
Here $M$ is the $2\times2$ transfer matrix for the full double-PT
potential $V_L(z)$, which is specified by $V_b$, $\alpha_{\rm PT}$,
and $L$. At fixed $m$, the two columns of $M$ are the right
asymptotic amplitude vectors $(A_+,B_+)^{\mathrm T}$ of the
solutions of $\psi''+[m^2-V_L(z)]\psi=0$ with left asymptotic
amplitudes $(A_-,B_-)=(1,0)$ and $(0,1)$, respectively.
The scattering coefficients are analytically continued in $m$
to formulate the resonance condition at complex frequencies.

With the time dependence $e^{-imt}$, the purely outgoing conditions
in Eq.~\eqref{qnm-boundary-condition} require $A_-=B_+=0$.
The left-outgoing Jost solution is normalized by setting
$(A_-,B_-)=(0,1)$. Its right incoming amplitude is then
$B_+=[M]_{22}$. We define the scalar Jost function $F$ as this
amplitude. Its zeros satisfy the right outgoing condition $B_+=0$
and give the quasinormal frequencies:
\begin{equation}
    \label{eq:dpt-jost-condition}
    \begin{aligned}
    F(m;V_b,\alpha_{\rm PT},L)
    &\equiv [M(m;V_b,\alpha_{\rm PT},L)]_{22},\\
    F\!\left(m_n^{\rm DPT};V_b,\alpha_{\rm PT},L\right)
    &=0,
    \qquad n=0,1,2,\ldots.
    \end{aligned}
\end{equation}
Here $m_n^{\rm DPT}=m_n^{\rm DPT}(V_b,\alpha_{\rm PT},L)$
denotes a root tracked continuously as the potential parameters vary,
and $n$ labels the selected QNM branch.
Equation~\eqref{eq:dpt-jost-condition} applies at all separations
and determines the frequencies implicitly for general $L$.
At $L=0$, the two components merge into a single exactly solvable
PT barrier.

Below, we construct analytic approximations in different separation
regimes to explain pole motion and provide initial parameter estimates.
In the well-separated applications, the joint and damping-weighted
fits use a cavity model constructed from exact single-barrier
reflection amplitudes. We check the fitted spectra by direct integration
with outgoing boundary conditions for the full DPT potential.
The less separated $s=3$ example uses the full DPT spectrum in the fit.
For a real potential, if $m$ is a resonance frequency,
then $-m^*$ is also a resonance frequency.
We label a pair with nonzero real parts by $m_{n,\pm}$, with
$m_{n,-}=-m_{n,+}^{*}$, where $*$ denotes complex conjugation.
We mainly display the branch with $\operatorname{Re}m>0$.

\subsection{Low-Overtone Spectral Structure at Different Separation Scales}
\label{subsec:low-overtone}

At fixed low overtone number, we examine the roots of
Eq.~\eqref{eq:dpt-jost-condition} in three regimes:
\begin{equation}
    \ell\ll1,
    \qquad
    \ell=\mathcal O(1),
    \qquad
    \ell\gg1.
    \label{eq:dpt-low-regime-classification}
\end{equation}
These correspond to strongly overlapping barriers,
the transition from a single peak to two peaks,
and well-separated barriers, respectively.
This classification uses the component width as the geometric
reference scale.
For higher overtones, the frequency-dependent scale $|m|^{-1}$
must also be considered, as discussed in
Sec.~\ref{subsec:high-overtone}.

\subsubsection{Overlapping barriers:
\texorpdfstring{$\ell\ll1$}
{separation much smaller than the barrier width}}

At zero separation, the two components merge into a single PT barrier
with twice the component height:
\begin{equation}
    V_0(z)=2V_b\operatorname{sech}^{2}(\alpha_{\rm PT}z).
\end{equation}
Its quasinormal frequencies are known exactly 
\cite{PoschlTeller:1933,Ferrari:1984pt}:
\begin{equation}
    m_{n,\pm}^{(0)}
    =\pm\sqrt{2V_b-\frac{\alpha_{\rm PT}^{2}}{4}}
    -i\alpha_{\rm PT}\left(n+\frac12\right),
    \qquad n=0,1,2,\ldots.
    \label{eq:dpt-l0-spectrum}
\end{equation}
The oscillatory branch requires $2V_b>\alpha_{\rm PT}^2/4$,
or equivalently $V_b/\alpha_{\rm PT}^2>1/8$.
This threshold refers to the merged barrier of height $2V_b$
and differs from the isolated-component threshold $1/4$.
The shallow-barrier applications below use direct integration for the full
potential at finite $L$ and do not require a nonzero real part
in the zero-separation reference spectrum.

Equation~\eqref{eq:dpt-l0-spectrum} explains the vertical pole sequence
at $L=0$ in Fig.~\ref{fig:double-barrier-spectral-features}:
on the branch with positive real part, all poles have the same real
frequency, while adjacent damping rates differ by $\alpha_{\rm PT}$.

Set $x=\alpha_{\rm PT}z$ and $g(x)=\operatorname{sech}^{2}x$.
At fixed $V_b$ and $\alpha_{\rm PT}$, the symmetric component shifts
$x\pm\ell/2$ make the potential an even function of $\ell$.
For $\ell\ll1$, the normalized potential has the expansion
\begin{equation}
    \frac{V_L(z)}{V_b}
    =g\left(x+\frac{\ell}{2}\right)
    +g\left(x-\frac{\ell}{2}\right)
    =2g(x)+\frac{\ell^2}{4}g''(x)
    +\mathcal O(\ell^4).
    \label{eq:dpt-even-taylor-step}
\end{equation}
Using $g''(x)=4g(x)-6g^2(x)$ gives
\begin{align}
    V_L(z)
    &=2V_b\operatorname{sech}^{2}(\alpha_{\rm PT}z)
    +\ell^2\Delta V_2(z)
    +V_b\,\mathcal O(\ell^4),
    \notag\\
    \Delta V_2(z)
    &=\frac{V_b}{2}
    \left[
    2\operatorname{sech}^{2}(\alpha_{\rm PT}z)
    -3\operatorname{sech}^{4}(\alpha_{\rm PT}z)
    \right].
    \label{eq:dpt-small-potential}
\end{align}
The dimensionless Jost function expands as
$F=F_0(m)+\ell^2F_2(m)+\mathcal O(\ell^4)$
at fixed $m$, $V_b$, and $\alpha_{\rm PT}$.
For an isolated simple zero $m_n^{(0)}$ of $F_0$, the frequency
expansion at fixed $n$ gives
\begin{equation}
    m_n(L)
    =m_n^{(0)}+\ell^2\delta m_n^{(2)}
    +\alpha_{\rm PT}\,\mathcal O(\ell^4),
    \qquad
    \delta m_n^{(2)}
    =-\frac{F_2(m_n^{(0)})}
    {\partial_mF_0(m_n^{(0)})}.
    \label{eq:dpt-small-jost-shift}
\end{equation}
A bilinear identity for the unperturbed outgoing state
\cite{Leung:1994perturbation,Zworski:2017resonances}
gives the same coefficient
\begin{equation}
    \delta m_n^{(2)}
    =
    \frac{
    \displaystyle\int_{\mathcal C}
    \Delta V_2(z)[\psi_n^{(0)}(z)]^2dz}
    {
    \displaystyle2m_n^{(0)}
    \int_{\mathcal C}[\psi_n^{(0)}(z)]^2dz}.
    \label{eq:dpt-small-bilinear-shift}
\end{equation}
This bilinear pairing contains no complex conjugation.
The integrals can be defined by analytic continuation from a domain
of convergence; equivalently, one may choose an exterior
complex-scaling contour $\mathcal C$ along which the outgoing
states decay \cite{Leung:1994perturbation,Zworski:2017resonances}.

The required integral ratios can be evaluated without constructing
the wave functions explicitly.
We take the branch with positive real part and write the unperturbed
outgoing state as $\psi_n$ inside the integrals.
Let $\eta_0=\sqrt{2V_b/\alpha_{\rm PT}^2-1/4}$
and define $\mathcal J_{r,n} =\int_{\mathcal C}\psi_n^2 \operatorname{sech}^{2r}(\alpha_{\rm PT}z)\,dz$.
Differentiating the unperturbed eigenvalue $(m_n^{(0)})^2$
with respect to the merged barrier height $2V_b$,
and using an integration-by-parts identity with
$\tanh(\alpha_{\rm PT}z)$ as a multiplier, gives
\begin{equation}
    \frac{\mathcal J_{1,n}}{\mathcal J_{0,n}}
    =\frac{m_n^{(0)}}{\alpha_{\rm PT}\eta_0},
    \qquad
    \frac{\mathcal J_{2,n}}{\mathcal J_{1,n}}
    =
    \frac{
    2[(m_n^{(0)})^2+\alpha_{\rm PT}^2+2V_b]}
    {8V_b+3\alpha_{\rm PT}^2}.
\end{equation}
The boundary terms vanish on the chosen complex-scaled contour.
These identities can also be obtained by analytic continuation
of convergent PT integrals.
Substitution into Eq.~\eqref{eq:dpt-small-bilinear-shift}
then gives the quadratic coefficient in closed form:
\begin{equation}
    \delta m_{n,+}^{(2)}
    =
    \frac{V_b}{2\alpha_{\rm PT}\eta_0}
    \frac{
    2V_b-3(m_{n,+}^{(0)})^2}
    {8V_b+3\alpha_{\rm PT}^2}.
    \label{eq:dpt-small-general-shift}
\end{equation}
The coefficient $\delta m_n^{(2)}$ multiplies the dimensionless
$\ell^2$ and has the same units as $m$.

The $\operatorname{sech}^{4}$ term in
Eq.~\eqref{eq:dpt-small-potential} produces an overtone-dependent
frequency shift.
Its effect on the entire QNM sequence cannot generally be reproduced
by a common adjustment of the height and width of a single PT barrier.

For example, when $V_b=\alpha_{\rm PT}^{2}$,
Eq.~\eqref{eq:dpt-small-general-shift} gives
\begin{align}
    \frac{m_0}{\alpha_{\rm PT}}
    &=\frac{\sqrt7}{2}-\frac{i}{2}
    +\ell^2\left(-\frac{5\sqrt7}{154}+\frac{21i}{154}\right)
    +\mathcal O(\ell^4),
    \notag\\
    \frac{m_1}{\alpha_{\rm PT}}
    &=\frac{\sqrt7}{2}-\frac{3i}{2}
    +\ell^2\left(\frac{7\sqrt7}{154}+\frac{63i}{154}\right)
    +\mathcal O(\ell^4),
    \notag\\
    \frac{m_2}{\alpha_{\rm PT}}
    &=\frac{\sqrt7}{2}-\frac{5i}{2}
    +\ell^2\left(\frac{31\sqrt7}{154}+\frac{105i}{154}\right)
    +\mathcal O(\ell^4).
    \label{eq:dpt-unit-example-small-spectrum}
\end{align}
The fundamental mode initially moves toward a lower real frequency
and a smaller damping rate, whereas the next two modes initially move
toward higher real frequencies and smaller damping rates.
Thus, even at small separation, the direction of the real-frequency
shift depends on the overtone number.

Table~\ref{tab:dpt-small-example} compares the quadratic predictions
with numerical results from the Bernstein polynomial spectral method
\cite{Fortuna:2023bernstein}.
Unless otherwise stated, the numerical examples comparing the three
separation regimes use $V_b=\alpha_{\rm PT}^{2}$ and units
in which $\alpha_{\rm PT}=1$.
The plot labels give $L$ in units of $\alpha_{\rm PT}^{-1}$,
so the reported values are those of $\ell$.
At $\ell=0.5$, the absolute complex-frequency error is of order
$10^{-3}\alpha_{\rm PT}$ for the fundamental mode and grows with $n$.

\begin{table}[htb]
    \caption{
    Comparison of the quadratic predictions
    $m_n^{\rm quad}=m_n^{(0)}+\ell^2\delta m_n^{(2)}$
    with the numerical frequencies $m_n^{\rm num}$ at $\ell=0.5$
    and $V_b=\alpha_{\rm PT}^{2}$.
    The accuracy of the fixed-overtone expansion decreases as
    the overtone number increases for these modes.
    }
    \begin{ruledtabular}
    \begin{tabular}{cccc}
        $n$
        & $m_n^{\rm num}/\alpha_{\rm PT}$
        & $m_n^{\rm quad}/\alpha_{\rm PT}$
        & $|m_n^{\rm num}-m_n^{\rm quad}|/\alpha_{\rm PT}$ \\
        0
        & $1.30229-0.46749i$
        & $1.30140-0.46591i$
        & $1.81\times10^{-3}$ \\
        1
        & $1.35072-1.40641i$
        & $1.35294-1.39773i$
        & $8.96\times10^{-3}$ \\
        2
        & $1.43733-2.35446i$
        & $1.45602-2.32955i$
        & $3.12\times10^{-2}$
    \end{tabular}
    \end{ruledtabular}
    \label{tab:dpt-small-example}
\end{table}

Additional direct-integration tests were performed for
$V_b/\alpha_{\rm PT}^2=1,3$ and $n=0,3,4$.
For $n=4$, reducing $\ell$ from $0.1$ to $0.05$
decreases the quadratic-approximation error from
$4.41\times10^{-4}$ to $2.82\times10^{-5}$
when $V_b/\alpha_{\rm PT}^2=1$,
and from $2.12\times10^{-4}$ to $1.34\times10^{-5}$
when $V_b/\alpha_{\rm PT}^2=3$,
all in units of $\alpha_{\rm PT}$.
These reductions are close to a factor of sixteen, as expected
when the quartic term in $\ell$ dominates the error.
The corresponding fundamental-mode errors are smaller.
The complete results and variations obtained with a second
integration contour are included in the accompanying data;
the contour-induced frequency changes are below
$2\times10^{-11}\alpha_{\rm PT}$.

\begin{figure*}[htb]
    \centering
    \includegraphics[width=0.95\textwidth]{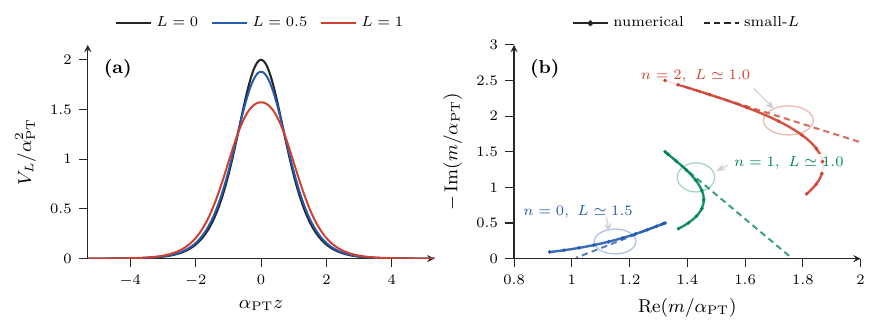}
    \caption{
    Potential profiles and low-overtone evolution in the overlapping
    regime and toward the formation of two peaks,
    with $V_b=\alpha_{\rm PT}^{2}$ and $\alpha_{\rm PT}=1$.
    (a) Exact potential profiles at $\ell=0,0.5,1$.
    (b) Trajectories of the first three QNMs computed by direct integration
    with exterior complex scaling over $0\le\ell\le3$
    (solid lines with filled markers), compared with the quadratic
    small-separation predictions of
    Eq.~\eqref{eq:dpt-unit-example-small-spectrum} (dashed lines).
    The dashed curves are formally continued beyond the quantitative
    range of validity of the quadratic expansion.
    Colored ellipses mark the onset of visible deviations:
    $\ell\simeq1.5$ for $n=0$
    and $\ell\simeq1$ for $n=1,2$.
    Light-gray arrows connect the highlighted regions
    to their corresponding labels.
    }
    \label{fig:dpt-low-overlapping}
\end{figure*}

Figure~\ref{fig:dpt-low-overlapping} further compares the quadratic
expansion with numerical trajectories for the full potential.
The quadratic predictions depart progressively from the numerical
results as the separation and overtone number increase.

The growth of the error with $n$ reflects the nonuniformity
of the fixed-overtone expansion at high overtone number.
For $V_b=\alpha_{\rm PT}^{2}$,
Eq.~\eqref{eq:dpt-small-general-shift} gives
\begin{align}
    \frac{\delta m_{n,+}^{(2)}}{\alpha_{\rm PT}}
    &=
    \frac{\sqrt7}{154}[6n(n+1)-5]
    +\frac{3i}{11}\left(n+\frac12\right),
    \notag\\
    \frac{m_{n+1,+}-m_{n,+}}{\alpha_{\rm PT}}
    &=-i
    +\ell^2\left[
    \frac{6\sqrt7}{77}(n+1)+\frac{3i}{11}
    \right]
    +\mathcal O(\ell^4).
    \label{eq:dpt-small-large-n-warning}
\end{align}
The spacing in Eq.~\eqref{eq:dpt-small-large-n-warning}
is expanded as $\ell\to0$ at fixed $n$, and the remainder
may depend on $n$.
The magnitude $|m_n^{(0)}|/\alpha_{\rm PT}$ grows linearly with $n$,
while the real part of $\delta m_{n,+}^{(2)}/\alpha_{\rm PT}$
grows quadratically.
Their ratio in the frequency expansion gives the necessary condition
$n\ell^2\ll1$ for the quadratic correction to remain perturbative
relative to the unperturbed complex frequency.

The displacement of the complex poles introduces the further scale
$n\ell$. When $n\ell\gtrsim1$, the characteristic scale associated
with the frequency is small enough to resolve the two shifted
complex-pole neighborhoods, and higher-order displacement terms
may become equally important.
The small-separation expansion at fixed $n$ and the high-overtone
analysis at fixed $\ell>0$ must therefore be treated separately.
Their relation is discussed further in
Sec.~\ref{subsec:high-overtone}.

The perturbative formulas above assume a nonzero, isolated simple pole.
The Jost-function expression requires
$\partial_mF_0(m_n^{(0)})\neq0$,
while the bilinear integral expression also requires
$\mathcal J_{0,n}\neq0$.
Multiple roots or exceptional points require a degenerate analysis
and may lead to Puiseux expansions.
No such degeneracy was encountered along the three numerical branches
tracked in Fig.~\ref{fig:dpt-low-overlapping}.

\subsubsection{Transition region:
\texorpdfstring{$\ell=\mathcal O(1)$}
{separation comparable to the barrier width}}

The critical separation at which the central potential changes shape
can be obtained exactly.
Let $s_L=\operatorname{sech}^{2}(\ell/2)$.
Then
\begin{equation}
    V_L''(0)
    =4V_b\alpha_{\rm PT}^{2}s_L(2-3s_L),
    \qquad
    L_c=\frac{2}{\alpha_{\rm PT}}
    \operatorname{arcosh}\sqrt{\frac32}
    \simeq\frac{1.317}{\alpha_{\rm PT}}.
    \label{eq:dpt-critical-separation}
\end{equation}
For $L<L_c$, the origin is a maximum of the full potential;
at $L=L_c$, the central curvature vanishes;
and for $L>L_c$, the origin becomes a minimum between two peaks.
This geometric transition does not itself require a nonsmooth change
in the quasinormal frequencies:
a root can vary smoothly with $L$ as long as it remains simple.
As the two peaks develop, cavity propagation and reflection phases
become increasingly important for understanding pole motion.

For $L>L_c$, the right-hand peak position $z_{\rm pk}^{\rm DPT}$,
its height $H_{\rm pk}$, and its curvature-defined inverse width
$\beta_{\rm pk}$ can be obtained analytically.
Writing $\tau_L=\tanh^2(\ell/2)$, we obtain
\begin{equation}
    \begin{aligned}
    z_{\rm pk}^{\rm DPT}
    &=\frac{1}{\alpha_{\rm PT}}
    \operatorname{arctanh}
    \sqrt{\frac{3\tau_L-1}{\tau_L(3-\tau_L)}},\\
    H_{\rm pk}
    &=V_L(z_{\rm pk}^{\rm DPT})
    =V_b\frac{(1+\tau_L)^2}{4\tau_L},\\
    \beta_{\rm pk}^2
    &=\alpha_{\rm PT}^{2}
    \frac{(3\tau_L-1)(3-\tau_L)}{4\tau_L}.
    \end{aligned}
    \label{eq:dpt-local-peak-parameters}
\end{equation}
At fixed $\alpha_{\rm PT}$ and $L\to\infty$,
$z_{\rm pk}^{\rm DPT}-L/2\to0$, $H_{\rm pk}\to V_b$, and $\beta_{\rm pk}\to\alpha_{\rm PT}$.
Just above $L_c$, however, $\beta_{\rm pk}$ is small,
the curvature-defined local width is large,
and overlap still has a substantial effect on the full potential.
The existence of two peaks therefore does not imply that the barriers
can already be treated independently.

Table~\ref{tab:dpt-local-peak-geometry} shows how the local peak
parameters approach those of the isolated components.
For example, at $\alpha_{\rm PT}=1$ and $\ell=1.5$,
the origin is already a minimum,
but both peaks remain close to the center and are substantially
broader than the isolated component barriers.

\begin{table}[htb]
    \caption{
    Right-hand peak parameters from
    Eq.~\eqref{eq:dpt-local-peak-parameters},
    with $V_b=\alpha_{\rm PT}^{2}$ and $\alpha_{\rm PT}=1$.
    }
    \begin{ruledtabular}
    \begin{tabular}{cccc}
        $\ell$ & $z_{\rm pk}^{\rm DPT}$ & $H_{\rm pk}$ & $\beta_{\rm pk}$ \\
        1.5 & 0.482 & 1.221 & 0.582 \\
        2.0 & 0.920 & 1.076 & 0.879 \\
        2.5 & 1.222 & 1.027 & 0.958 \\
        3.0 & 1.490 & 1.010 & 0.985
    \end{tabular}
    \end{ruledtabular}
    \label{tab:dpt-local-peak-geometry}
\end{table}

Reflection symmetry reduces the spectral problem for the full potential
exactly to parity conditions on the half-line.
Let $f_+(z,m,L)$ be the right-outgoing Jost solution:
\begin{equation}
    f_+''+[m^2-V_L(z)]f_+=0,
    \qquad
    f_+(z,m,L)\sim e^{imz}
    \quad(z\to+\infty).
\end{equation}
For $m\neq0$, its value and first derivative at $z=0$ can be written as
\begin{align}
    f_+(0)&=A_c+B_c,
    \qquad
    f_+'(0)=im(A_c-B_c),
    \notag\\
    A_c&=\frac{imf_+(0)+f_+'(0)}{2im},
    \qquad
    B_c=\frac{imf_+(0)-f_+'(0)}{2im}.
    \label{eq:dpt-center-projection}
\end{align}
This is an algebraic decomposition of the central Cauchy data
in a local free-wave basis and does not require $V_L(0)=0$.
We then define the central reflection function
$\mathcal R_c=B_c/A_c$:
\begin{equation}
    \mathcal R_c(m,L)
    =
    \frac{imf_+(0)-f_+'(0)}
    {imf_+(0)+f_+'(0)}.
    \label{eq:dpt-center-reflection}
\end{equation}
The function $\mathcal R_c$ contains the effect of the full potential
on the right half-line, including the overlapping tail
of the left component.
The full-line QNM conditions become
\begin{equation}
    \mathcal R_c(m,L)
    =
    \begin{cases}
        +1, & f_+'(0)=0
        \quad\text{(even parity)},\\
        -1, & f_+(0)=0
        \quad\text{(odd parity)}.
    \end{cases}
    \label{eq:dpt-parity-conditions}
\end{equation}
If $m_n(L)$ is a simple root of fixed parity,
its variation with separation satisfies
\begin{equation}
    \frac{dm_n}{dL}
    =
    -\frac{
    \partial_L\mathcal R_c(m_n,L)}
    {\partial_m\mathcal R_c(m_n,L)}.
    \label{eq:dpt-middle-migration}
\end{equation}
This implicit differential relation applies throughout the transition
region as long as the root remains simple.
Changes in mode labels caused by reordering the poles according
to damping rate must be distinguished from the actual motion
of a continuously tracked branch.

For weakly damped QNMs, these conditions also admit a phase interpretation.
At a real frequency where $\mathcal R_c$ is finite and nonzero,
choose a locally analytic branch of its logarithm.
On the real-frequency axis, write $\mathcal R_c=|\mathcal R_c|e^{i\Phi_c}$
and define $a_c=\ln|\mathcal R_c|$.
For $m=m_R-i\gamma$, expanding as $\gamma/\alpha_{\rm PT}\to0$
at fixed $m_R$ and fixed potential parameters gives
\begin{equation}
    \ln\mathcal R_c(m_R-i\gamma,L)
    =
    a_c+\gamma\partial_m\Phi_c
    +i\left(\Phi_c-\gamma\partial_ma_c\right)
    +\mathcal O\!\left((\gamma/\alpha_{\rm PT})^2\right).
    \label{eq:dpt-middle-log-expansion}
\end{equation}
The real-axis functions and their derivatives are evaluated at $m=m_R$.
Matching this expression to $ij\pi$ with the phase integer $j$
yields the coupled first-order relations
\begin{equation}
    a_c+\gamma\partial_m\Phi_c=0,
    \qquad
    \Phi_c-\gamma\partial_ma_c=j\pi.
    \label{eq:dpt-middle-coupled-phase}
\end{equation}
For weak damping, with $\partial_m\Phi_c\neq0$ and
$|\partial_ma_c|\ll|\partial_m\Phi_c|$,
the leading-order relations are
\begin{equation}
    \Phi_c(m_R,L)=j\pi,
    \qquad
    \gamma\simeq
    -\frac{\ln|\mathcal R_c(m_R,L)|}
    {\partial_m\Phi_c(m_R,L)},
    \qquad j\in\mathbb Z.
    \label{eq:dpt-middle-phase-damping}
\end{equation}
Even and odd values of $j$ correspond to even and odd parity,
respectively.
When the reflection magnitude varies rapidly with frequency,
the amplitude-derivative term in
Eq.~\eqref{eq:dpt-middle-coupled-phase} must be retained.

The derivative $\partial_m\Phi_c$ is the effective phase delay
of the reflection process measured at the center.
As the cavity expands, the propagation contribution to this delay
increases.
If the reflection loss does not increase correspondingly,
the damping rate decreases.
This relation explains the motion of resonance poles toward
the real axis under the stated conditions.

To construct a semianalytic approximation, we describe
the right-hand peak by a local PT barrier with the parameters
in Eq.~\eqref{eq:dpt-local-peak-parameters}.
For $H\operatorname{sech}^{2}(\beta z)$,
the reflection amplitude is \cite{Bueno:2018echoes}
\begin{equation}
    R_{\rm PT}(m;H,\beta)
    =
    \frac{\Gamma(im/\beta)}{\Gamma(-im/\beta)}
    \frac{
    \Gamma(1-im/\beta-\lambda)
    \Gamma(\lambda-im/\beta)}
    {\Gamma(1-\lambda)\Gamma(\lambda)},
    \qquad
    \lambda(1-\lambda)=\frac{H}{\beta^2}.
    \label{eq:dpt-pt-reflection}
\end{equation}
Approximating propagation from the center to the reference peak
as free propagation then gives
\begin{equation}
    \mathcal R_c^{\rm med}(m,L)
    \simeq
    R_{\rm PT}(m;H_{\rm pk},\beta_{\rm pk})e^{2iS_c(m,L)},
    \qquad
    S_c(m,L)=mz_{\rm pk}^{\rm DPT}.
    \label{eq:dpt-middle-pt-approximation}
\end{equation}
Here $R_{\rm PT}$ uses the isolated-barrier center as its phase reference,
so translation to $z_{\rm pk}^{\rm DPT}$ introduces the factor $e^{2imz_{\rm pk}^{\rm DPT}}$.
The reflection phase is unwrapped continuously from the positive
real-frequency axis, and the same logarithmic branch is used
when searching for complex roots.

In this local PT approximation, $S_c$ describes the free-propagation
phase, while $R_{\rm PT}$ contains the local reflection phase and leakage.
The effect of the full central potential remains encoded in the exact
reflection function in Eq.~\eqref{eq:dpt-center-reflection}.
Since the PT reflection amplitude already includes the scattering
phase accumulated within the barrier, introducing an additional
action through the barrier requires a consistent matching prescription
to avoid double counting phase contributions.

Writing $R_{\rm PT}=|R_{\rm PT}|e^{i\delta_{\rm PT}}$,
we define the total phase
\begin{equation}
    \Theta(m,L)
    =2S_c(m,L)+\delta_{\rm PT}(m;H_{\rm pk},\beta_{\rm pk}).
    \label{eq:dpt-middle-total-phase}
\end{equation}
With this phase convention, even $j$ corresponds to even parity
and odd $j$ to odd parity.
For weak damping, if the amplitude-derivative correction to the
phase is negligible and $\partial_m\Theta\neq0$, we obtain
\begin{equation}
    \Theta(m_R,L)=j\pi,
    \qquad
    \gamma\simeq
    -\frac{
    \ln|R_{\rm PT}(m_R;H_{\rm pk},\beta_{\rm pk})|}
    {\partial_m\Theta(m_R,L)}.
    \label{eq:dpt-middle-pt-phase-damping}
\end{equation}
At fixed phase quantum number $j$, the real frequency also satisfies
\begin{equation}
    \frac{dm_R}{dL}
    =
    -\frac{
    2\partial_LS_c+\partial_L\delta_{\rm PT}}
    {2\partial_mS_c+\partial_m\delta_{\rm PT}}.
    \label{eq:dpt-middle-real-migration}
\end{equation}
The denominator is the effective phase delay in this approximate model,
while the numerator contains the separation dependence of both
the propagation and reflection phases.
These contributions can have opposite signs,
allowing some overtones to move initially toward higher real frequencies
and then turn toward lower real frequencies as the cavity-propagation
contribution grows.

The numerical branches for $V_b=\alpha_{\rm PT}^{2}$
and $\alpha_{\rm PT}=1$ exhibit this behavior.
Table~\ref{tab:dpt-transition-trajectories} lists two branches
obtained by numerical continuation from the $n=1,2$ modes at $L=0$.
The damping rates of both branches decrease over the listed separations.
The $n=1$ branch turns toward lower real frequency between
$\ell=2$ and $2.5$, whereas the $n=2$ branch traces a broader arc.
These changes reflect the actual motion of continuous branches.

\begin{table}[htb]
    \caption{
    Representative QNM frequencies obtained by direct integration for
    $V_b=\alpha_{\rm PT}^{2}$ and $\alpha_{\rm PT}=1$.
    Each entry gives $(\operatorname{Re}m,-\operatorname{Im}m)$.
    Branch labels follow the QNMs continued from $L=0$.
    }
    \begin{ruledtabular}
    \begin{tabular}{ccc}
        $\ell$ & Initial branch $n=1$ & Initial branch $n=2$ \\
        1.5 & $(1.450,0.958)$ & $(1.798,1.733)$ \\
        2.0 & $(1.454,0.742)$ & $(1.864,1.421)$ \\
        2.5 & $(1.423,0.563)$ & $(1.861,1.143)$ \\
        3.0 & $(1.370,0.421)$ & $(1.813,0.909)$
    \end{tabular}
    \end{ruledtabular}
    \label{tab:dpt-transition-trajectories}
\end{table}

For branches continued from $L=0$, we assess how well the local
PT approximation satisfies the central parity condition using the residual
\begin{equation}
    r_n(L)
    =
    \left|
    \mathcal R_c^{\rm med}(m_n^{\rm num},L)-\mathcal P_n
    \right|,
    \qquad \mathcal P_n=(-1)^n.
    \label{eq:dpt-middle-residual}
\end{equation}
For $V_b=\alpha_{\rm PT}^{2}$ and $\alpha_{\rm PT}=1$,
the free-propagation approximation $S_c=mz_{\rm pk}^{\rm DPT}$ gives
$(r_0,r_1,r_2)=(1.513,0.783,1.168)$ at $\ell=1.5$
and $(0.0379,0.0185,0.0298)$ at $\ell=3$.
The local PT approximation therefore has a substantial mismatch
just above $L_c$, with improved agreement as the components
separate over the range examined.
These quantities are dimensionless residuals of the parity condition;
the corresponding frequency errors are reported below.
The useful separation range depends on the potential parameters.

\begin{figure*}[htb]
    \centering
    \includegraphics[width=0.95\textwidth]{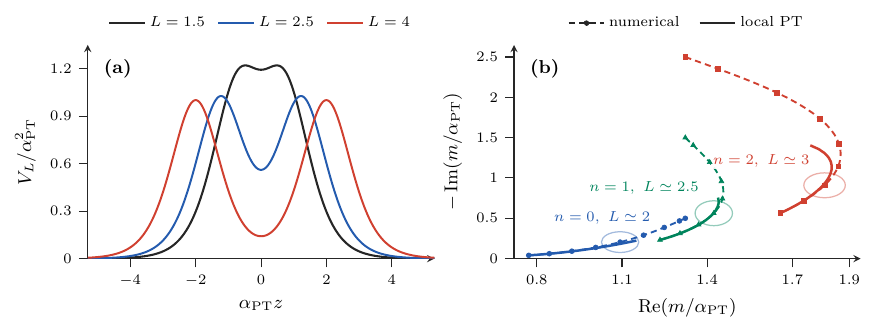}
    \caption{
    Potential profiles and spectral evolution in the transition region,
    with $V_b=\alpha_{\rm PT}^{2}$ and $\alpha_{\rm PT}=1$.
    (a) Exact potential profiles at $\ell=1.5,2.5,4$.
    (b) Numerical trajectories over $0\le\ell\le4$
    (dashed lines with filled markers), compared with roots of
    the local PT approximation
    $\mathcal R_c^{\rm med}=\pm1$
    over $2\le\ell\le4$ (solid lines).
    The latter use $(z_{\rm pk}^{\rm DPT},H_{\rm pk},\beta_{\rm pk})$ from
    Eq.~\eqref{eq:dpt-local-peak-parameters}
    and the free-propagation phase $S_c=mz_{\rm pk}^{\rm DPT}$ from
    Eq.~\eqref{eq:dpt-middle-pt-approximation}.
    Circles, triangles, and squares denote branches continued
    from $n=0,1,2$, respectively.
    Small colored ellipses highlight representative samples
    at $\ell=2,2.5,3$ on the $n=0,1,2$ branches, respectively.
    }
    \label{fig:dpt-low-transition}
\end{figure*}

Figure~\ref{fig:dpt-low-transition} compares the local PT approximation
with QNM frequencies obtained by direct integration for the full potential.
At $\ell=3$, the absolute differences for the first three quasinormal
frequencies are $7.1\times10^{-3}$, $3.5\times10^{-3}$,
and $6.6\times10^{-3}$, in units where $\alpha_{\rm PT}=1$.

\subsubsection{Separated barriers:
\texorpdfstring{$\ell\gg1$}
{separation much larger than the barrier width}}

For QNMs at fixed low overtone number, well-separated component
barriers can be approximated as independent scatterers
\cite{Mostafazadeh:2020transfer}.
Using their respective centers as phase references, we write
the transfer matrix as
\begin{equation}
    M_{\rm tot}(m,L)
    =M_R(m)P(m,L)M_L(m)+\delta M(m,L),
    \qquad
    P(m,L)=
    \begin{pmatrix}
        e^{imL}&0\\
        0&e^{-imL}
    \end{pmatrix}.
    \label{eq:dpt-large-transfer-factorization}
\end{equation}
Here $M_{\rm tot}$ is expressed in local plane-wave bases
referenced to the two component centers.
It differs from the global plane-wave transfer matrix in
Eq.~\eqref{eq:dpt-transfer-matrix} by diagonal phase factors.
These multiply the $22$ entry by a nonvanishing factor
and therefore leave the resonance condition unchanged.
The correction $\delta M$ accounts for barrier overlap.
The exponential scale in Eq.~\eqref{eq:dpt-midpoint-overlap}
characterizes the potential overlap itself.

Neglecting $\delta M$, repeated cavity round trips generate
the geometric series
$1+R_b^2e^{2imL}+R_b^4e^{4imL}+\cdots$,
leading to the round-trip resonance condition
\begin{equation}
    1-e^{2imL}R_b^2(m)=0,
    \qquad
    R_b(m)=R_{\rm PT}(m;V_b,\alpha_{\rm PT}).
    \label{eq:dpt-large-cavity-condition}
\end{equation}
For nonzero $m$ with finite $R_b(m)$, this is the symmetric form of the distant-barrier scattering relation
\cite{Cheung:2022flea,Ianniccari:2024deciphering}.
Taking a square root and substituting the exact single-barrier
reflection amplitude in Eq.~\eqref{eq:dpt-pt-reflection},
with consistent branch and phase conventions,
gives the implicit Gamma-function resonance condition
for the separated double-PT model used in
Refs.~\cite{Voelkel:2020thesis,Bueno:2018echoes}.
We use this established relation to examine the spectral structure
and accuracy of the independent-barrier approximation.

For $V_b>\alpha_{\rm PT}^{2}/4$, let
\begin{equation}
    \lambda_b=\frac12+i\eta,
    \qquad
    \eta=\sqrt{\frac{V_b}{\alpha_{\rm PT}^{2}}-\frac14}.
    \label{eq:dpt-component-shape}
\end{equation}
The poles of an isolated component barrier, indexed by overtone number $n$, are
\begin{equation}
    \Omega_{n,\pm}
    =\pm\alpha_{\rm PT}\eta
    -i\alpha_{\rm PT}\left(n+\frac12\right).
    \label{eq:dpt-single-barrier-poles}
\end{equation}
Choose one of the simple poles $\Omega_{n,\pm}$, denote it by $\Omega_n$,
and write $R_b(m)\simeq\rho_n/(m-\Omega_n)$ near this pole,
where $\rho_n$ is its residue.
Substitution into Eq.~\eqref{eq:dpt-large-cavity-condition} gives
\begin{equation}
    (m-\Omega_n)^2
    \simeq\rho_n^2e^{2imL}.
    \label{eq:dpt-near-pole-quadratic}
\end{equation}
Evaluating the propagation factor initially at $m=\Omega_n$
gives the first iteration for the local root splitting,
\begin{equation}
    m_{n,\pm}^{\rm split}
    \simeq\Omega_n\pm\rho_n e^{i\Omega_nL}.
    \label{eq:dpt-pole-migration}
\end{equation}
Here $\pm$ labels two local branches near the same isolated pole.
This approximation requires the frequency shift to remain within
the domain of the single-pole expansion and to satisfy
$|\rho_n e^{i\Omega_nL}L|\ll1$.
Since
$|e^{i\Omega_nL}|=e^{-\operatorname{Im}(\Omega_n)L}$
grows exponentially with $L$, a distant barrier need not constitute
a small spectral perturbation.
When the local expansion conditions fail, the roots must be obtained
from Eq.~\eqref{eq:dpt-large-cavity-condition}.
Thus, even a structure far from the original barrier can substantially
shift damped resonance frequencies.

Away from zeros and poles of $R_b$, taking a locally analytic
logarithm of Eq.~\eqref{eq:dpt-large-cavity-condition} gives the
cavity sequence
\begin{align}
    m_N
    &=\frac{\pi N}{L}
    +\frac{i}{L}\operatorname{Log}R_b(m_N),
    \notag\\
    \operatorname{Re}m_N
    &\simeq
    \frac{\pi N-\phi_b(\pi N/L)}{L},
    \qquad
    \operatorname{Im}m_N
    \simeq
    \frac{1}{L}
    \ln\left|R_b\left(\frac{\pi N}{L}\right)\right|.
    \label{eq:dpt-large-cavity-roots}
\end{align}
On the positive real axis, write $R_b=|R_b|e^{i\phi_b}$ with
the continuous reflection phase fixed by $\phi_b(0^+)=-\pi$.
We continue this logarithmic branch to complex frequencies.
Here $N$ is the cavity phase quantum number.
When the frequency dependence of the reflection phase can be neglected,
the real-frequency spacing is approximately $\pi/L$.
The damping rate is set by the logarithmic amplitude loss per round trip
divided by the round-trip time.
Equation~\eqref{eq:dpt-large-cavity-roots} gives
$R_b(m_N)e^{im_NL}=(-1)^N$, so even $N$ gives even parity and odd $N$
gives odd parity.

Retaining the frequency dependence of the reflection amplitude
gives a phase-delay correction.
For $m=m_R-i\gamma$ with weak damping, let $a_b=\ln|R_b|$ and expand
the locally analytic logarithm about $m_R$.
The first-order expansion of the cavity condition gives
\begin{align}
    m_RL+\phi_b(m_R)
    -\gamma\partial_ma_b(m_R)
    &=\pi N,
    \notag\\
    \gamma\left[L+\partial_m\phi_b(m_R)\right]
    &=-a_b(m_R).
    \label{eq:dpt-large-phase-delay}
\end{align}
If the amplitude-derivative term in the first line is subleading
and $L+\partial_m\phi_b\neq0$, we obtain
\begin{equation}
    \Delta m_R
    \simeq\frac{\pi}{L+\partial_m\phi_b},
    \qquad
    \gamma
    \simeq-\frac{\ln|R_b|}{L+\partial_m\phi_b}.
    \label{eq:dpt-large-effective-length}
\end{equation}
The effective optical length in this approximation therefore includes
both the geometric separation and the Wigner reflection delay
\cite{Wigner:1955delay}.
Equation~\eqref{eq:dpt-large-effective-length} is also the large-separation
form of the central phase relation
in Eq.~\eqref{eq:dpt-middle-phase-damping},
connecting the phase descriptions in the transition
and separated-barrier regimes.

\begin{figure*}[htb]
    \centering
    \includegraphics[width=0.95\textwidth]{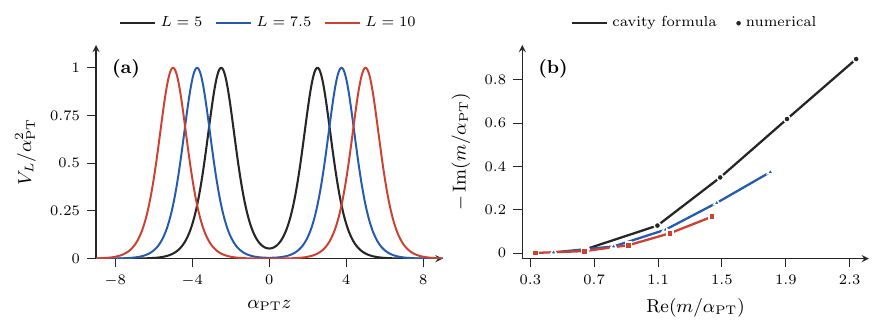}
    \caption{
    Potential profiles and low-overtone spectra in the separated-barrier
    regime, with $V_b=\alpha_{\rm PT}^{2}$ and $\alpha_{\rm PT}=1$.
    (a) Exact potential profiles at $\ell=5,7.5,10$.
    (b) The first five cavity roots at each fixed separation,
    ordered from lower left to upper right as $n=0,\ldots,4$.
    With the reflection-phase convention above, these roots have $N=n$.
    Colored solid lines connect roots obtained independently from
    Eq.~\eqref{eq:dpt-large-cavity-condition};
    filled markers show the corresponding numerical roots
    of the full double-PT potential.
    The lines only connect roots of the approximate model.
    Black circles, blue triangles, and red squares correspond
    to $\ell=5,7.5,10$, respectively.
    }
    \label{fig:dpt-low-separated}
\end{figure*}

Figure~\ref{fig:dpt-low-separated} compares the separated-barrier
cavity condition with numerical roots of the full potential
at three fixed separations, $\ell=5,7.5,10$.

The maximum absolute complex-frequency error among the five roots
decreases from $1.8\times10^{-4}$ at $\ell=5$
to $2.2\times10^{-6}$ at $\ell=7.5$
and $2.2\times10^{-8}$ at $\ell=10$.
This behavior is consistent with exponential suppression of barrier overlap;
the three separation samples give an effective decay trend
of approximately $e^{-1.8\ell}$.
This trend describes the finite parameter range examined
and cannot be promoted directly to a uniform bound on resonance-root errors.
All three separation regimes are based on the full Jost condition,
with each providing an analytic approximation for understanding
the separation dependence of the QNM spectrum.

\subsection{High-Frequency Scattering Model and Finite-Overtone Tests}
\label{subsec:high-overtone}

We first derive the high-frequency spectral relation of the
independent-barrier scattering model and explain how its complex
optical length is related to the nearest component-barrier poles.
We then test its applicability over finite overtone ranges
using numerical spectra of the full summed potential.
For the full potential, the global asymptotic problem still requires
an analysis of Stokes connections and branch selection.

\subsubsection{Distinct small-separation and high-overtone limits}

High overtones introduce a second characteristic scale, $|m|^{-1}$.
With $V_b$ and $\alpha_{\rm PT}$ fixed, the small-separation
expansion takes $\ell\to0$ at fixed $n$ and is nonuniform in $n$.
At fixed $\ell>0$, sufficiently large $|m|$ resolves the two
shifted complex-singularity neighborhoods. The two limits are
\begin{equation}
    \begin{array}{ll}
    \ell\to0,\quad n\ \text{fixed},
    & \text{small-separation expansion at fixed overtone number},
    \\[2pt]
    n\to\infty,\quad \ell>0\ \text{fixed},
    & \text{high-frequency analysis with resolved complex poles}.
    \end{array}
    \label{eq:dpt-distinct-limits}
\end{equation}
At $\ell=0$, the spectral problem reduces to a single barrier
of doubled height, with frequencies given by
Eq.~\eqref{eq:dpt-l0-spectrum}.
At fixed $\ell>0$, the high-frequency analysis considered here requires
\begin{equation}
    |m|\gg\alpha_{\rm PT},
    \qquad
    |m|L\gg1.
    \label{eq:dpt-resolution-condition}
\end{equation}
The first condition is the local high-frequency limit,
while the second resolves the horizontal displacement
between the two sets of complex singularities.
For $\ell\ll1$, the scale of the unperturbed spectrum gives
the crossover estimate $n_{\rm cross}\sim\ell^{-1}$.
Thus, the quadratic small-separation formula can accurately describe
the first few poles while failing at sufficiently high overtone number.

\subsubsection{Complex singularities and scattering quantization}

The component-barrier poles nearest the real axis are
\begin{equation}
    z_L^{\pm}
    =-\frac{L}{2}\pm\frac{i\pi}{2\alpha_{\rm PT}},
    \qquad
    z_R^{\pm}
    =+\frac{L}{2}\pm\frac{i\pi}{2\alpha_{\rm PT}}.
    \label{eq:dpt-nearest-complex-poles}
\end{equation}
Their local effects can be analyzed using the
Wentzel--Kramers--Brillouin (WKB) momentum $Q(z)$, defined by
$Q^2(z)=m^2-V_L(z)$ \cite{Berry:1972wkb,Andersson:2004stokes}.
For a component centered at $z_c=\pm L/2$, the expansion near one
of its poles $z_p$ at fixed $L>0$ is
\begin{equation}
    \operatorname{sech}^{2}
    [\alpha_{\rm PT}(z-z_c)]
    =
    -\frac{1}{\alpha_{\rm PT}^{2}(z-z_p)^2}
    +\mathcal O(1),
    \qquad
    Q^2(z)
    =
    m^2+\frac{V_b}{\alpha_{\rm PT}^{2}(z-z_p)^2}
    +V_b\,\mathcal O(1).
    \label{eq:dpt-local-pole-momentum}
\end{equation}
The corresponding pair of turning points is
\begin{equation}
    z_{p,\pm}
    =
    z_p\pm\frac{i\sqrt{V_b}}{\alpha_{\rm PT}m}
    +\alpha_{\rm PT}^{-1}\,\mathcal O\!\left((\alpha_{\rm PT}/m)^3\right),
    \qquad
    (L,V_b,\alpha_{\rm PT}\ \text{fixed},\ L>0).
    \label{eq:dpt-turning-points-near-pole}
\end{equation}
As $|m|\to\infty$, each turning-point pair contracts toward its pole.
Using the scaled local coordinate $\bar{z}=m(z-z_p)$ gives
the leading local equation
\[
    \frac{d^2\psi}{d\bar{z}^{2}}
    +\left(1+\frac{V_b}{\alpha_{\rm PT}^2\bar{z}^{2}}\right)\psi=0.
\]
Writing $\psi=\sqrt{\bar{z}}\,f_{\rm B}(\bar{z})$ yields a Bessel equation
of order $\nu_{\rm B}$, with $\nu_{\rm B}^{2}=1/4-V_b/\alpha_{\rm PT}^2$.
Thus, a finite local connection coefficient can survive
even as the turning-point neighborhood shrinks.
The pole locations provide geometric information about possible connections,
while the global Stokes structure also depends on the full potential
and the chosen branch.

For the branch with $\operatorname{Re}m>0$
and $\operatorname{Im}m<0$,
a diagonal connection compatible with the candidate spectral ray below
involves the neighborhoods of $z_L^-$ and $z_R^+$.
Their displacement is
\begin{equation}
    \mathcal L
    =z_R^+-z_L^-
    =L+\frac{i\pi}{\alpha_{\rm PT}}.
    \label{eq:dpt-complex-length}
\end{equation}
Its real part is the separation between the component barrier centers,
and its imaginary part is the vertical distance between the selected
lower and upper PT poles.
Along the candidate spectral direction,
$\operatorname{Im}(m\mathcal L)$ remains of order unity,
while $\operatorname{Re}(m\mathcal L)$ grows as $\pi n$.
This complex length therefore controls the high-frequency spectral scale
of the scattering model.

This interpretation draws on one-dimensional complex-WKB
connection analysis
\cite{Berry:1972wkb,Andersson:2004stokes}.
Related complex-plane monodromy methods have also been used
to study highly damped black-hole QNMs \cite{Motl:2003monodromy}.
For a fixed branch of $Q(z)=\sqrt{m^2-V_L(z)}$ and a turning point
$z_{\rm tp}$, define $S(z)=\int_{z_{\rm tp}}^zQ(z')\,dz'$.
The local anti-Stokes curves issuing from $z_{\rm tp}$ satisfy
$\operatorname{Im}S(z)=0$.
The two WKB exponentials $e^{\pm iS(z)}$ then have equal magnitudes.
More generally, along a level curve $\operatorname{Im}S(z)=C$,
their magnitude ratio is the constant
$|e^{iS}|/|e^{-iS}|=e^{-2C}$.
With the chosen phase normalization, $C=0$ corresponds
to strictly equal magnitudes.

Using $Q\simeq m$ outside the pole neighborhoods,
the diagonal displacement is compatible with the leading
equal-magnitude direction of the candidate spectrum:
$\operatorname{Im}(m\mathcal L)=\mathcal O(1)$,
whereas $\operatorname{Im}(mL)=\mathcal O(n)$.
This comparison tests the compatibility of the proposed connection
with the spectral direction.
Determining the dominant connection selected by outgoing conditions
still requires tracing the full Stokes graph from both asymptotic ends
on a fixed branch of $Q$ and comparing competing paths.
Local matching also contributes a finite imaginary part to the action,
so the straight segment between the two poles cannot be identified
directly with an exact anti-Stokes curve.

For the oscillatory branch of an isolated barrier,
$V_b>\alpha_{\rm PT}^{2}/4$, use the component parameter
$\lambda_b$ from Eq.~\eqref{eq:dpt-component-shape}.
The high-frequency limit of the exact single-barrier reflection amplitude gives
\begin{equation}
    \rho_{\lambda_b}=-2i\sin(\pi\lambda_b),
    \qquad
    R_b(m)\sim
    \rho_{\lambda_b}e^{-\pi m/\alpha_{\rm PT}}.
\end{equation}
Specifically, we first apply the Gamma-function reflection identity
to Eq.~\eqref{eq:dpt-pt-reflection}
and then use a Stirling expansion in the sector containing
the candidate QNM ray.
The exponential factor reflects the distance from the real axis
to the nearest PT pole, while $\rho_{\lambda_b}$ gives
the local connection coefficient.

The round-trip condition of the independent-barrier model is
$1-R_b^2e^{2imL}=0$.
Combining the two reflection factors with free propagation gives
\begin{equation}
    e^{-2im(L+i\pi/\alpha_{\rm PT})}
    =\rho_{\lambda_b}^{2}
    =-4\sin^2(\pi\lambda_b)
    =-4\cosh^2(\pi\eta).
    \label{eq:dpt-high-master-condition}
\end{equation}
This quantization condition combines single-barrier reflection
amplitudes with free propagation.
For the full summed potential, overlap and competing global
connections can modify the resulting spectrum.

With the same $\eta$ from Eq.~\eqref{eq:dpt-component-shape}, define
\begin{equation}
    \chi=\ln\left[4\cosh^2(\pi\eta)\right].
\end{equation}
The leading logarithmic condition is
\begin{equation}
    -2im\mathcal L
    =\chi-i(2j_{\log}+1)\pi,
    \qquad j_{\log}\in\mathbb Z.
    \label{eq:dpt-high-logarithmic-branch}
\end{equation}
Large positive $j_{\log}$ gives the high-frequency branch with
positive real and negative imaginary parts.
For a selected sequence of consecutive overtones, write
$j_{\log}=n+j_0$, with a fixed integer $j_0$.
At fixed $V_b$, $\alpha_{\rm PT}$, and $L>0$,
the high-frequency sequence of the independent-barrier model
has the expansion
\begin{equation}
    m_n=
    \frac{\pi(n+j_0+\frac12)+i\chi/2}
    {L+i\pi/\alpha_{\rm PT}}
    +\alpha_{\rm PT}\,\mathcal O(n^{-1}),
    \qquad n\to\infty.
    \label{eq:dpt-high-spectrum}
\end{equation}
The remainder follows from the subleading reflection-amplitude
expansion given below.
The expansion applies in the sector specified there,
with $|m_n|\gg\alpha_{\rm PT}$ and $|m_n|L\gg1$.

The integer $j_0$ sets the numbering origin along the ray
and leaves its leading spacing and transverse offset unchanged.
These quantities provide the spectral descriptors used below
to construct initial parameter estimates.
Their accuracy for the full DPT potential is assessed
over the tested overtone ranges.
The parameter estimates are subsequently checked against
selected QNM frequencies of the full potential.

Subleading terms in the single-barrier reflection amplitude
can also be expanded systematically.
Let $u=-im/\alpha_{\rm PT}$.
Using the Gamma-function reflection identity,
Eq.~\eqref{eq:dpt-pt-reflection} can be rewritten as
\begin{equation}
    R_b(m)
    =
    -\frac{\sin(\pi\lambda_b)}{\sin(\pi u)}
    \frac{
    \Gamma(u+1-\lambda_b)\Gamma(u+\lambda_b)}
    {u\,\Gamma^2(u)}.
    \label{eq:dpt-reflection-rearranged}
\end{equation}
The Gamma-function factor has the expansion
$1-\lambda_b(1-\lambda_b)/u+\mathcal O(u^{-2})$,
while for $\operatorname{Im}u<0$,
$1/\sin(\pi u)\sim2i e^{-i\pi u}$.
These expansions apply in the corresponding sector away
from the negative real $u$ axis, with
$\operatorname{Re}m/\alpha_{\rm PT}\to\infty$.
Using $\lambda_b(1-\lambda_b)=V_b/\alpha_{\rm PT}^{2}$ gives
\begin{equation}
    R_b(m)
    =
    \rho_{\lambda_b} e^{-\pi m/\alpha_{\rm PT}}
    \left[
    1-i\frac{V_b}{\alpha_{\rm PT}m}
    +\mathcal O\left(\frac{\alpha_{\rm PT}^{2}}{m^2}\right)
    +\mathcal O\left(e^{-2\pi m/\alpha_{\rm PT}}\right)
    \right].
    \label{eq:dpt-reflection-subleading}
\end{equation}
Substituting this expansion into the independent-barrier resonance
condition gives a correction of order $\alpha_{\rm PT}/n$
to the displayed high-frequency sequence.
This estimate concerns the high-frequency expansion
of the single-barrier amplitudes within that model.

Extending the result to the full summed potential requires
additional control of overlap corrections, propagation through
the central potential, and competing global connections.
A bound uniform in both $n$ and $\ell$
has not been established.
Consequently, exponential suppression of the potential overlap
at fixed frequency does not by itself justify interchanging
the large-separation and high-overtone limits.

We therefore test Eq.~\eqref{eq:dpt-high-spectrum} as a candidate
description of selected sequences in the full-potential spectrum.
Establishing its high-overtone asymptotic validity for the full
potential requires a global connection analysis on a fixed branch
of $Q$.

\subsubsection{Numerical tests for the full double-PT potential}

We compute QNMs of the full summed potential by direct integration,
imposing outgoing boundary conditions and central parity conditions,
to test Eq.~\eqref{eq:dpt-high-spectrum}.
Initial guesses for even- and odd-parity roots are chosen
near the candidate branch, so this calculation tests the selected sequence.

\begin{figure}[htb]
\centering
\includegraphics[width=0.95\linewidth]{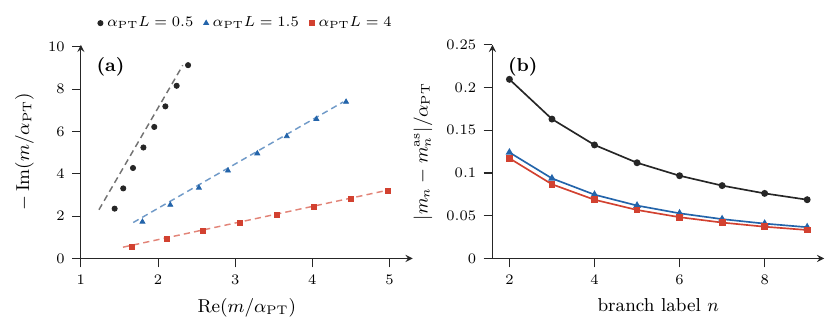}
\caption{
High-overtone tests for the full double-PT potential.
(a) Numerical poles (markers) and the high-frequency approximation
(dashed lines) for $n=2,\ldots,9$, with alternating even and odd parity.
(b) Absolute complex-frequency differences between the numerical
frequencies $m_n$ and $m_n^{\rm as}$, the leading approximation in
Eq.~\eqref{eq:dpt-high-spectrum} evaluated with $j_0=0$.
We take $V_b=\alpha_{\rm PT}^{2}$ and use units
in which $\alpha_{\rm PT}=1$.
}
\label{fig:dpt-high-numerics}
\end{figure}

For $V_b=\alpha_{\rm PT}^2$,
Fig.~\ref{fig:dpt-high-numerics} shows that the difference
between numerical frequencies and scattering predictions decreases
as $n=2,\ldots,9$ increases at
$\ell=0.5,1.5,4$.
These separations cover the overlapping, transition,
and separated-barrier regimes.
The absolute differences between half the spacing of the two
highest odd-parity frequencies shown and the predicted increment
$\pi/(L+i\pi/\alpha_{\rm PT})$
are $0.00829$, $0.00475$, and $0.00437$, respectively,
in units of $\alpha_{\rm PT}$.
Section~\ref{subsec:numerical-method} describes the numerical method,
parity assignments, and contour tests.
These results support the scattering relation as a description
of the selected finite-overtone sequence.
Its global asymptotic properties and other branches
require further analysis.

The initial parameters in the braneworld applications below include
the shallow-barrier branch $0<V_b<\alpha_{\rm PT}^2/4$.
The full potential must therefore also be tested when its components
lie on this branch.
We use the two sets of initial parameters in
Sec.~\ref{subsec:reconstruction}
and recompute the selected QNMs of the full summed potential by direct integration.
Table~\ref{tab:shallow-test} compares slopes fitted to five consecutive
quasinormal frequencies with the analytically continued
scattering prediction.
As the selected windows move to higher overtones,
the relative difference decreases from approximately
$3.0\times10^{-4}$ to $3.9\times10^{-5}$,
while the minimum $|m|/\alpha_{\rm PT}$ within each window
increases from $1.6$--$1.7$ to $4.6$--$5.1$.
The largest frequency change between two integration configurations
is $5.2\times10^{-7}k$.
These results provide a direct numerical test on the shallow-barrier
branch, restricted to the finite overtone ranges examined.

\begin{table}[htb]
\caption{
Tests of the full DPT potential with shallow component barriers,
using the initial parameters for $s=19,21$ in
Sec.~\ref{subsec:reconstruction}.
Frequencies and slopes are given in units of $k$,
and $n$ labels the DPT sequence.
The last column gives
$|\Delta m_{\rm fit}-\pi/(L+i\pi/\alpha_{\rm PT})|
/|\pi/(L+i\pi/\alpha_{\rm PT})|$,
the relative difference between the fitted window slope
and the scattering prediction.
}
\label{tab:shallow-test}
\begin{ruledtabular}
\begin{tabular}{cccc}
$s$ & $n$ window & $\Delta m_{\rm fit}/k$ & Relative difference \\
19 & 20--24 & $1.566446-0.132081i$ & $3.03\times10^{-4}$ \\
19 & 40--44 & $1.566788-0.132064i$ & $8.44\times10^{-5}$ \\
19 & 60--64 & $1.566860-0.132062i$ & $3.90\times10^{-5}$ \\
21 & 20--24 & $1.568190-0.119444i$ & $3.02\times10^{-4}$ \\
21 & 40--44 & $1.568531-0.119423i$ & $8.46\times10^{-5}$ \\
21 & 60--64 & $1.568603-0.119421i$ & $3.90\times10^{-5}$ \\
\end{tabular}
\end{ruledtabular}
\end{table}

\subsubsection{Parameter information from high-frequency spectral rays}

We now examine the parameter information contained in the
independent-barrier high-frequency scattering model.
Let $N_n=\pi(n+j_0+1/2)$ and use
$|\mathcal L|^2=L^2+\pi^2/\alpha_{\rm PT}^{2}$.
Separating the real and imaginary parts of
Eq.~\eqref{eq:dpt-high-spectrum} gives the leading expressions
\begin{align}
    \operatorname{Re}m_n
    &\sim
    \frac{N_nL+\chi\pi/(2\alpha_{\rm PT})}{|\mathcal L|^2},
    \notag\\
    -\operatorname{Im}m_n
    &\sim
    \frac{N_n\pi/\alpha_{\rm PT}-\chi L/2}{|\mathcal L|^2}.
    \label{eq:dpt-high-components}
\end{align}
Within this model, the limiting adjacent complex-frequency spacing is
\begin{equation}
    \Delta m_{\infty}
    \equiv\lim_{n\to\infty}(m_{n+1}-m_n)
    =\frac{\pi}{L+i\pi/\alpha_{\rm PT}}
    =\frac{\pi L}{|\mathcal L|^2}
    -i\frac{\pi^2/\alpha_{\rm PT}}{|\mathcal L|^2},
    \qquad
    \frac{-\operatorname{Im}m_n}{\operatorname{Re}m_n}
    \to\frac{\pi}{\ell}.
    \label{eq:dpt-high-spacing}
\end{equation}
Writing $\Delta m_\infty=\Delta_R-i\Delta_I$ gives
\begin{equation}
    \frac{\Delta_I}{\Delta_R}
    =\frac{\pi}{\ell},
    \qquad
    |\Delta m_\infty|
    =\frac{\pi}{\sqrt{L^2+\pi^2/\alpha_{\rm PT}^{2}}}.
    \label{eq:dpt-ray-analysiss}
\end{equation}
The ray angle therefore determines the ratio of the separation
to the component width, while the spacing magnitude supplies
the corresponding dimensional scale.
In practice, these quantities can be extracted by complex linear fits
to consecutive overtones.
Varying the selected overtone range tests how finite-frequency
corrections affect the fitted slope.

The strength parameter $\chi$ controls the transverse offset
of the spectral ray, whereas $L$ and $\alpha_{\rm PT}$
control the leading complex spacing.
These relations are based on the high-frequency scattering model
at fixed $L>0$.
The spectrum at $L=0$ is instead given by
Eq.~\eqref{eq:dpt-l0-spectrum}
and must be treated as a distinct limit.

When Eq.~\eqref{eq:dpt-high-spectrum} describes the selected sequence,
a measured spacing $\Delta m=\Delta_R-i\Delta_I$,
with $\Delta_R,\Delta_I>0$,
allows Eq.~\eqref{eq:dpt-high-spacing} to be inverted
without prior knowledge of $V_b$:
\begin{equation}
    L=\frac{\pi\Delta_R}{\Delta_R^2+\Delta_I^2},
    \qquad
    \alpha_{\rm PT}
    =\frac{\Delta_R^2+\Delta_I^2}{\Delta_I}.
    \label{eq:dpt-spacing-inversion-preview}
\end{equation}
Once these two scales are fixed, the transverse offset gives
the strength parameter and a height estimate:
\begin{equation}
    \chi\simeq
    2\operatorname{Im}\left[
    m_n\left(L+\frac{i\pi}{\alpha_{\rm PT}}\right)
    \right],
    \qquad
    \eta=\frac{1}{\pi}
    \operatorname{arcosh}\left(\frac{e^{\chi/2}}{2}\right),
    \qquad
    V_b=\alpha_{\rm PT}^{2}\left(\eta^2+\frac14\right).
    \label{eq:dpt-height-inversion-preview}
\end{equation}
The oscillatory branch of the isolated barrier,
$V_b>\alpha_{\rm PT}^{2}/4$,
corresponds to $\chi>\ln4$.

The same spacing relation can be continued to the shallow-barrier branch.
For $0<V_b<\alpha_{\rm PT}^{2}/4$,
choose $\lambda_b\in(1/2,1)$ such that
$\lambda_b(1-\lambda_b)=V_b/\alpha_{\rm PT}^{2}$.
The local multiplier in Eq.~\eqref{eq:dpt-high-master-condition}
is then $\rho_{\lambda_b}^2=-4\sin^2(\pi\lambda_b)$, giving
\begin{equation}
    \chi=\ln\left[4\sin^2(\pi\lambda_b)\right],
    \qquad
    \lambda_b
    =1-\frac{1}{\pi}
    \arcsin\left(\frac{e^{\chi/2}}{2}\right),
    \qquad
    V_b=\alpha_{\rm PT}^{2}\lambda_b(1-\lambda_b).
    \label{eq:dpt-height-inversion-shallow}
\end{equation}
This branch satisfies $\chi<\ln4$,
with equality at the common threshold.
This branch gives $\lambda_b\to1$ as $V_b\to0$.
The complementary value $1-\lambda_b$ gives the same height
and scattering amplitude.

For $V_b/\alpha_{\rm PT}^2<1/8$,
the $L=0$ reference spectrum no longer has a nonzero oscillation frequency.
The shallow-barrier high-frequency relation remains based on
the scattering model at fixed $L>0$
and belongs to a different limit from the spectral problem
obtained by setting $L=0$ first.

At fixed $\alpha_{\rm PT}$, the ideal transverse-offset relation
determines a unique height within the corresponding branch.
The conditions $\chi<\ln4$ and $\chi>\ln4$
select the shallow-barrier branch and the oscillatory branch
of the isolated barrier, respectively.
Finite-frequency corrections and uncertainty in the estimated $\chi$
affect both branch selection and height inference.
The stability of branch identification is particularly important
near the threshold.
Low-overtone QNMs provide complementary information by testing
the reference potential's predictions for trapping and damping,
allowing parameter refinement and independent validation.

Finally, the leading scattering relation also reveals parameter
nonuniqueness in asymmetric models.
For two isolated barriers on their oscillatory branches,
with parameters $(V_j,\alpha_j)$, $j=1,2$,
the corresponding replacements are
\begin{equation}
    \mathcal L
    \to
    L+\frac{i\pi}{2}
    \left(\frac1{\alpha_1}+\frac1{\alpha_2}\right),
    \qquad
    \chi
    \to
    \ln\left[
    4\cosh(\pi\eta_1)\cosh(\pi\eta_2)
    \right],
    \label{eq:dpt-asymmetric-generalization}
\end{equation}
where $\eta_j=\sqrt{V_j/\alpha_j^2-1/4}$.
The ray spacing and transverse offset then constrain combinations
of the two widths and strengths rather than the parameters
of each barrier separately.
Further constraints require low-overtone data
or other independent scattering information.
Within the scattering description adopted here,
low-overtone QNMs test the reference potential's predictions
for trapping and damping, while spacings between higher overtones
provide initial estimates of optical scales.
Combining these complementary constraints forms the basis
of the parameter-inference procedure developed in the next section.

\section{Finite-Spectrum Constraints within a Reference Family}
\label{sec:spectral-inversion}

This section examines the constraints that finite sets of quasinormal
frequencies place on model parameters within a chosen reference family
and assesses the predictive accuracy of the reference potential
for different spectral features.
We construct initial parameter estimates from analytic scattering
relations, refine them through spectral fitting, and assess the resulting
constraints using QNM frequencies computed by direct integration for the full
potential and predictions for QNMs excluded from the fit.

\subsection{Parameter Definitions and Initial Estimates}
\label{subsec:inversion-target}
\label{subsec:inversion-high}

For the symmetric DPT reference potential, we use the component parameters
\begin{equation}
    \boldsymbol\theta=(V_b,\alpha_{\rm PT},L).
    \label{eq:inversion-parameters}
\end{equation}
Let $\boldsymbol\theta_{\rm fit}(\mathcal I,\boldsymbol w)$ denote the
parameters obtained by minimizing the fitting residual over the QNMs
with indices in $\mathcal I$, where $\boldsymbol w$ denotes the
prescribed weights.
Their values depend on the selected QNMs, weights, and objective function.
A finite set of quasinormal frequencies is insufficient to uniquely
determine an arbitrary effective potential $V(z)$
\cite{Zworski:2017resonances}.
We therefore discuss parameter constraints within an explicitly
specified reference family.

To interpret the fitted parameters, we introduce two sets of benchmarks
evaluated directly from the known background.
The first consists of the geometric parameters
$\boldsymbol\theta_{\rm geo}
=(V_{\rm pk},\alpha_{\rm loc},L_{\rm geo})$,
determined from the peak heights, peak curvatures, and peak separation
according to Eq.~\eqref{barrier-descriptors}.
For the analytic background family considered below,
we define the complex-pole reference parameters
$\boldsymbol\theta_{\rm pole}
=(V_{\rm pole},\alpha_{\rm pole},L_{\rm pole})$
by matching the locations and leading Laurent coefficients
of the nearest complex poles of the background potential
to those of the DPT reference potential.
Both benchmark sets are obtained without using the fitted QNM
frequencies and serve to interpret the spectral fits.

We denote the initial estimate from a finite spectral window $W$ by
$\boldsymbol\theta_{\rm sp}(W)$,
where $W$ is a set of two or more consecutive overtone indices.
This estimate is constructed from the spectral spacing and transverse
offset in the high-frequency scattering model.
Numerical comparisons below examine its relation to the geometric
parameters, complex-pole reference parameters, and final fitted parameters.
The correspondence between inferred parameters and potential geometry
is assessed separately from the spectral predictive accuracy
of the reference potential.

Low-overtone QNMs provide information about trapping and leakage,
while a consecutive sequence of higher-overtone QNMs provides an estimate
of the complex optical length within the scattering model.
Combining these two sources of information tests which spectral features
a single reference potential can reproduce simultaneously,
and with what accuracy.
For a wave equation with real coefficients, resonance roots
with nonzero real parts occur in pairs satisfying
$m_{n,-}=-m_{n,+}^*$.
We therefore mainly display the branch with positive real part.
This conjugation symmetry follows from the real coefficients
of the equation and does not require spatial reflection symmetry
of the potential.

For numerically reliable higher-overtone QNMs with consecutive overtone
numbers and unambiguous labels, a complex linear fit yields the
spacing estimate $\Delta m=\Delta_R-i\Delta_I$.
When $\Delta_R,\Delta_I>0$,
Eq.~\eqref{eq:dpt-spacing-inversion-preview} gives
\begin{equation}
    L_{\rm sp}
    =\frac{\pi\Delta_R}{\Delta_R^2+\Delta_I^2},
    \qquad
    \alpha_{\rm sp}
    =\frac{\Delta_R^2+\Delta_I^2}{\Delta_I}.
    \label{eq:inversion-spacing}
\end{equation}
These relations invert the parameter map for the asymptotic sequence
of the scattering model and provide initial estimates when applied
to finite-frequency windows.
Finite-frequency corrections can shift the estimates away
from the DPT parameters that generated the spectrum.
Calibrating this finite-window bias using DPT spectra with known
parameters therefore helps interpret subsequent differences
from the geometric barrier parameters.

For the transverse offset, we use the estimator
\begin{equation}
    \chi_{\rm sp}(W)
    \equiv
    \frac{2}{|W|}
    \sum_{n\in W}
    \operatorname{Im}\left[
    m_n\left(
    L_{\rm sp}+\frac{i\pi}{\alpha_{\rm sp}}
    \right)\right].
    \label{eq:inversion-intercept}
\end{equation}
The leading ray relation gives $\chi\simeq\chi_{\rm sp}(W)$.
The component height is then estimated using
Eq.~\eqref{eq:dpt-height-inversion-preview}
or Eq.~\eqref{eq:dpt-height-inversion-shallow},
depending on the branch.
The former applies to the oscillatory branch of an isolated barrier,
and the latter to the shallow-barrier branch.

The response of the initial estimates to changes in the fitted spacing
follows by linearizing Eq.~\eqref{eq:inversion-spacing}.
The relative changes obey
\begin{align}
    \frac{\delta L_{\rm sp}}{L_{\rm sp}}
    &=
    \frac{\Delta_I^2-\Delta_R^2}{\Delta_R^2+\Delta_I^2}
    \frac{\delta\Delta_R}{\Delta_R}
    -\frac{2\Delta_I^2}{\Delta_R^2+\Delta_I^2}
    \frac{\delta\Delta_I}{\Delta_I},
    \notag\\
    \frac{\delta\alpha_{\rm sp}}{\alpha_{\rm sp}}
    &=
    \frac{2\Delta_R^2}{\Delta_R^2+\Delta_I^2}
    \frac{\delta\Delta_R}{\Delta_R}
    +\frac{\Delta_I^2-\Delta_R^2}{\Delta_R^2+\Delta_I^2}
    \frac{\delta\Delta_I}{\Delta_I}.
    \label{eq:inversion-sensitivity}
\end{align}
For $\Delta_I/\Delta_R\ll1$, the two responses reduce to
$\delta L_{\rm sp}/L_{\rm sp}\simeq-\delta\Delta_R/\Delta_R$
and
$\delta\alpha_{\rm sp}/\alpha_{\rm sp}
\simeq2\delta\Delta_R/\Delta_R-\delta\Delta_I/\Delta_I$.
In the two $n=5,\ldots,9$ windows considered below,
$\Delta_I/\Delta_R\simeq0.084$ and $0.076$.
The absolute values of the coefficients multiplying $\delta\Delta_I/\Delta_I$
in the response of $L_{\rm sp}$ are approximately $0.014$ and $0.012$,
respectively.
Thus, in these windows, the initial separation estimate is controlled
mainly by the real part of the complex spacing,
while the inverse-width estimate remains sensitive
to relative changes in its imaginary part.

These linear relations characterize the local sensitivity
of the initialization map to the input spacing.
The geometric interpretation of the resulting parameters is tested
by comparison with the peak benchmarks,
while their frequency predictions are assessed using numerical spectra
of the full potential and QNMs excluded from the fit.

\subsection{Spectral Fitting and Accuracy Assessment}
\label{subsec:inversion-low}
\label{subsec:inversion-validation}

We use low-overtone QNMs to test how well the initial reference
potential describes trapping and leakage and to refine its parameters.
Let $r\in\{\mathrm{ov},\mathrm{tr},\mathrm{sep}\}$ denote
the overlapping, transition, and separated-barrier regimes
in Eq.~\eqref{eq:dpt-low-regime-classification}, respectively.
The corresponding spectral predictions
$m_n^{(r)}(\boldsymbol\theta)$ can be obtained from
the quadratic small-separation expansion in
Eq.~\eqref{eq:dpt-small-jost-shift} with the coefficient from
Eq.~\eqref{eq:dpt-small-general-shift},
the full parity condition~\eqref{eq:dpt-parity-conditions}
or its local PT approximation~\eqref{eq:dpt-middle-pt-approximation},
and the separated-barrier cavity condition
in Eq.~\eqref{eq:dpt-large-cavity-condition}, respectively.

Let $\mathcal I_{\rm fit}$ contain the indices of the input
frequencies $\widehat m_n$ selected for fitting. The weighted
sum of squared complex-frequency residuals is
\begin{equation}
    \mathcal E^{(r)}(\boldsymbol\theta;\iota)
    =
    \sum_{n\in\mathcal I_{\rm fit}}w_n
    \left|
    \widehat m_n
    -m_{\iota(n)}^{(r)}(\boldsymbol\theta)
    \right|^2,
    \qquad
    r\in\{\mathrm{ov},\mathrm{tr},\mathrm{sep}\}.
    \label{eq:low-overtone-fit-residual}
\end{equation}
Here $w_n>0$ specifies the weight of each QNM.
Any additional weights applied separately to real or imaginary parts
must be defined in the particular objective function.
The map $\iota$ specifies the correspondence between input frequencies
and spectral branches of the reference potential.
Once this correspondence is assigned,
the relevant roots are tracked continuously as the parameters vary,
with parity preserved when that information is available.
This prevents branch relabeling caused solely by changes
in the ordering of damping rates.

The full parity condition applies at all separations,
whereas the small-separation expansion, local PT approximation,
and separated-barrier condition each have their own validity scales.
The approximation regime must therefore be selected by considering
the fitting residual, the inferred $\ell$,
and the applicability conditions of the approximation.
Different low-overtone QNMs respond differently to changes in separation,
height, and width.
Joint fitting helps identify the spectral features associated
with these parameter variations.
Whether these frequencies supply independent parameter constraints
is examined quantitatively in Sec.~\ref{subsec:identifiability}.

A window $W$ used to extract spectral spacing should contain
numerically reliable QNMs with consecutive overtone numbers
and unambiguous labels.
In the following, we write $\alpha=\alpha_{\rm PT}$.
For each candidate parameter set, we first record
$\min_W|m|/\alpha$ and $\min_W|m|L$
to assess whether the selected frequencies reach the expected
high-frequency scales. The reflection-amplitude expansion also requires
$\operatorname{Re}m/\alpha\gg1$ in a sector where $u=-im/\alpha$
stays away from the negative real axis
[Eq.~\eqref{eq:dpt-reflection-subleading}].
Windows that do not satisfy these scale conditions may still
be used to construct initial estimates,
whose usefulness is determined by subsequent numerical validation.
We assess these estimates through finite-window bias calibration,
spectral calculations for the full potential,
and predictions for QNMs excluded from the fit.

After obtaining initial estimates, we refine the parameters
by fitting the selected QNM frequencies.
We then evaluate the frequency predictions using the full potential
and compare the results across different windows and against
QNMs excluded from the fit.
Let $W'$ be the comparison window and let $\theta_j(W)$ denote
the $j$th parameter estimated from window $W$. We use the same
estimation procedure and weights for both windows.
For a candidate parameter vector $\widehat{\boldsymbol{\theta}}$,
we define two dimensionless diagnostics for nonzero denominators.
\begin{equation}
    D_j
    =\frac{|\theta_j(W')-\theta_j(W)|}
    {|\theta_j(W)|},
    \qquad
    B_j
    =
    \left|
    \frac{
    \theta_j^{\rm ray}
    (m_W^{\rm syn}(\widehat{\boldsymbol{\theta}}))}
    {\widehat\theta_j}
    -1
    \right|.
    \label{eq:window-diagnostics}
\end{equation}
Here $\theta_j^{\rm ray}$ is the $j$th output of the ray map
defined above, and $m_W^{\rm syn}(\widehat{\boldsymbol{\theta}})$
is a synthetic spectrum generated at the candidate parameters in
window $W$, using the same branch labels. For the calibration in
Sec.~\ref{subsec:synthetic}, these frequencies are generated with the
cavity model constructed from exact single-barrier reflection amplitudes.

The quantity $D_j$ measures parameter changes induced by shifting
the window, while $B_j$ uses a synthetic DPT spectrum with known
parameters to calibrate the finite-window bias of the ray map
at that parameter point.
When parameter accuracy is reported directly from ray estimates,
both diagnostics should be compared with the corresponding
parameter tolerances.
For spectral fits initialized by ray estimates,
the assessment combines frequency residuals for the full potential,
window dependence, and predictions for QNMs excluded from the fit.
Absolute frequency tolerances and relative damping tolerances
should be specified separately.
The overtone range needed to reach a given accuracy depends
on the model parameters and the spectral features of interest.

For example, if the target accuracy for the inverse width is $1\%$,
the ray estimates from the $n=5$--$9$ windows considered below
do not meet this requirement because their synthetic biases
reach $4$--$5.4\%$.
Moreover, none of the six damping weights tested in
Sec.~\ref{subsec:damping} simultaneously yields
a root-mean-square (RMS) complex-frequency error below $0.05k$
for the QNMs excluded from the fit
and a relative error below $10\%$
in the fundamental-mode damping rate.
These results illustrate trade-offs between accuracy requirements
for the specified fitting setup;
the tolerances should be chosen according to the research objective.

The numerical analysis below uses established parity assignments
and QNM labels.
Alternative correspondences between frequencies and spectral branches
should be fitted separately and subjected to the same validation.

\subsection{Local Parameter Identifiability}
\label{subsec:identifiability}

To quantify local parameter constraints within the reference family,
we use a cavity spectral map constructed from exact single-barrier
reflection amplitudes and test this forward approximation
against selected QNM frequencies of the full DPT potential,
computed by direct integration.
For a selected QNM set $\mathcal I$,
define the dimensionless sensitivity matrix
\begin{equation}
    J_{aj}
    =\frac{\partial y_a}{\partial\log\vartheta_j},
    \qquad
    \boldsymbol y=
    (\operatorname{Re}m_n/k,\operatorname{Im}m_n/k)_{n\in\mathcal I},
    \qquad
    \vartheta=(kL,\alpha/k,V_b/k^2).
    \label{eq:log-jacobian}
\end{equation}
The model parameters have been reordered and made dimensionless.
In logarithmic parameter coordinates,
$J$ describes changes in dimensionless frequency components
induced by relative parameter variations.

Table~\ref{tab:jacobian} uses the representative equal-weight fit
at $k=1$,
$(L,\alpha,V_b)\simeq(2.00146,17.83048,13.27993)$.
All listed matrices have numerical rank three.
Thus, for this forward model and fixed branch labels,
the three parameters have locally independent spectral responses
at this parameter point.
Combining $n=0,1,2,5,6,7$ gives a smaller condition number
than using either $n=0$--$2$ or $n=5$--$9$ alone.

In the logarithmic parameter coordinates defined in
Eq.~\eqref{eq:log-jacobian},
the least-constrained direction for the joint QNM set
is proportional to $(-0.005,-0.852,-0.524)$.
This direction is dominated by combined changes in inverse width
and height, with only a small separation component,
indicating that the spectral responses to these two parameters
can partially compensate each other.
This analysis concerns the local sensitivity of joint spectral fitting.
It complements Eq.~\eqref{eq:inversion-sensitivity},
which describes the response of the initial parameter estimates
to changes in the complex spacing.

\begin{table}[htb]
\caption{
Singular values and condition numbers
$\kappa=\sigma_1/\sigma_3$
of the sensitivity matrix in Eq.~\eqref{eq:log-jacobian},
using dimensionless frequency components and logarithmic
parameter coordinates.
The entries use equal weights for all dimensionless frequency components.
Different parameter coordinates, component scalings, or weighting
schemes would give different condition numbers.
}
\label{tab:jacobian}
\begin{ruledtabular}
\begin{tabular}{ccccc}
Overtones $n$ & $\sigma_1$ & $\sigma_2$ & $\sigma_3$ & $\kappa$ \\
0--2 & 4.6642 & 1.1432 & 0.03341 & 139.6 \\
5--9 & 26.6433 & 1.1682 & 0.39022 & 68.3 \\
0,1,2,5,6,7 & 18.3347 & 1.4126 & 0.44031 & 41.6 \\
20--24 & 78.5324 & 4.7530 & 0.10615 & 739.9 \\
\end{tabular}
\end{ruledtabular}
\end{table}

Increasing the overtone number does not guarantee improved constraints
on every parameter.
At this parameter point, the $n=20$--$24$ window gives
a condition number of approximately $740$.
Its least-constrained direction in the same logarithmic parameter
coordinates is approximately $(-0.0019,-0.2425,0.9702)$
and is dominated by the height component.
Low-overtone QNMs or additional scattering data can provide
complementary information on the height,
while higher-overtone QNMs are used to estimate optical scales.

Doubling the finite-difference step used to compute $J$
from $10^{-4}$ to $2\times10^{-4}$ changes the resulting matrix
by less than $6\times10^{-9}$ in relative Frobenius norm.
We also perform six noiseless synthetic tests:
one of $L,\alpha,V_b$ is varied by $\pm20\%$
while the other two are held fixed.
Generating and fitting synthetic frequencies within the same
reflection-spectrum model and with fixed branch assignments
recovers all three parameters,
with relative differences below $10^{-10}$
between the fitted and generating parameters.
At each parameter point, the frequencies at $n=0,3,9$
are also checked by direct integration for the full DPT potential
to assess the accuracy of the forward approximation.

The sensitivity analysis quantifies local parameter constraints
within the specified reference model,
while the noiseless synthetic tests check the internal consistency
of parameter recovery.
Error assessment in applications must also include uncertainty
in the input frequencies, alternative frequency-to-branch assignments,
and mismatch between the reference potential
and the braneworld effective potential.

\section{Spectral Constraints and Systematic Errors in a Braneworld Family}
\label{sec:applications}

We use the analytic family of double-brane backgrounds in
Ref.~\cite{Tan:2024dbl}, where $s$ is a positive odd integer.
With $k=\delta=1$, the tensor dual potential is
\begin{equation}
    V_s^{\rm dual}(z)
    =
    \frac{3z^{2s-2}(z^{2s}+4s-2)}
    {4(1+z^{2s})^2}.
    \label{eq:double-brane-dual-potential}
\end{equation}
This potential is symmetric about $z=0$ and develops two peaks
for $s>1$.
Direct integration along complex contours uses the analytic
continuation of this rational function.
The complex-pole analysis below uses the same continuation.

We first obtain initial parameter estimates from spectral-ray fits
to five published Bernstein frequencies for each of $s=19,21$.
Additional spectral calculations, comparisons between input methods,
and variations of the fitting weights and potential profiles
then test the physical interpretation and predictive accuracy
of these estimates.
Comparisons between different values of $s$ are performed within
the same background family to examine how changes in internal structure
affect the spectral constraints.

Since $V_s^{\rm dual}(-z)=V_s^{\rm dual}(z)$,
the two barriers have equal peak heights and curvatures.
We therefore adopt the symmetric three-parameter DPT potential
as the reference model.
The parameter comparisons and accuracy assessments below
refer to this reflection-symmetric background family.
An asymmetric two-component PT reference potential has five
independent parameters,
$(V_1,V_2,\alpha_1,\alpha_2,L)$,
after fixing the overall translation.
The leading ray relation constrains combinations of the two widths
and strengths [Eq.~\eqref{eq:dpt-asymmetric-generalization}].
For asymmetric potentials, low-overtone QNMs can be computed by
two-sided direct integration with outgoing boundary conditions
without parity reduction.

\subsection{Finite-Window Estimates and Real-Axis Peak Geometry}
\label{subsec:reconstruction}
\label{subsec:comparison}

We use the sixth through tenth entries of the Bernstein column
in Table~3 of Ref.~\cite{Tan:2024dbl}, labeled $n=5,\ldots,9$ here.
We fit the complex frequencies against $n$ with equal weights
and average the transverse offset over $W=\{5,\ldots,9\}$.
The slope and offset estimate are
\begin{equation}
    \Delta m_{\rm fit}
    =
    \frac{\sum_{n=5}^{9}(n-7)m_n}{10},
    \qquad
    \chi_{\rm sp}(W)
    =
    \frac25\sum_{n=5}^{9}
    \operatorname{Im}\!\left[
    m_n\left(
    L_{\rm sp}+\frac{i\pi}{\alpha_{\rm sp}}
    \right)\right].
    \label{eq:brane-spectral-line-fit}
\end{equation}
The resulting fitted complex spacings are
\begin{equation}
    \frac{\Delta m^{(19)}_{\rm fit}}{k}
    =1.566921-0.132061i,
    \qquad
    \frac{\Delta m^{(21)}_{\rm fit}}{k}
    =1.568664-0.119420i.
    \label{eq:brane-fitted-spacings}
\end{equation}
Equations~\eqref{eq:inversion-spacing}
and~\eqref{eq:dpt-height-inversion-shallow}
give the initial parameters in
Table~\ref{tab:brane-spectral-reconstruction},
with transverse offsets
$\chi_{\rm sp}(W)=-2.678,-2.883$.
Both component heights satisfy
$0<V_b^{\rm sp}<\alpha_{\rm sp}^2/4$,
so the shallow-barrier height map is used.

We quantify the residual of the spectral-ray approximation by
\[
    \varepsilon_{\rm ray}
    =
    \left[
    \frac{1}{|W|}
    \sum_{n\in W}
    |m_n^{\rm ray}-m_n|^2
    \right]^{1/2},
    \qquad
    W=\{5,\ldots,9\}.
\]
Here $m_n^{\rm ray}$ is obtained from the leading expression in
Eq.~\eqref{eq:dpt-high-spectrum}, evaluated at the ray estimates
with $j_0=0$ to align the asymptotic sequence with the input overtone
indices.

The two parameter sets satisfy
$\alpha_{\rm sp}L_{\rm sp}=37.27,41.27$,
for which the PT components have very little overlap on the real axis.
However, the frequency scales within the windows are
$|m_n|/\alpha_{\rm sp}=0.463$--$0.797$
and $0.418$--$0.720$, respectively,
and do not satisfy the conditions of the local high-frequency expansion.
Table~\ref{tab:brane-spectral-reconstruction} therefore compares
the finite-window ray estimates
$\boldsymbol\theta_{\rm sp}(W)$
with geometric parameters
$\boldsymbol\theta_{\rm geo}$
determined directly from the real-axis peaks.
Their agreement is a numerical result for the selected windows.
The next subsection compares higher-overtone windows obtained by
direct integration.

\begin{table*}[htb]
    \caption{
    Finite-window spectral estimates and real-axis peak parameters
    for the dual potential of the double-brane background.
    Input frequencies are taken from the Bernstein column
    of Table~3 in Ref.~\cite{Tan:2024dbl},
    using $n=5,\ldots,9$ in the numbering adopted here.
    The superscript or subscript ``sp'' denotes ray estimates;
    $L_{\rm geo}$ is the real-axis peak separation,
    $\alpha_{\rm loc}$ is defined by the peak curvature,
    and $V_{\rm pk}$ is the peak height of the full potential.
    The table reports numerical values of $kL$, $\alpha/k$,
    $V/k^2$, and $\varepsilon_{\rm ray}/k$.
    Direct comparison between component-center separation
    and actual peak separation applies to the well-separated
    barriers considered here.
    $\varepsilon_{\rm ray}$ is the RMS residual of the analytic
    scattering sequence; prediction errors for the full DPT spectrum
    are assessed separately.
    Differences between the spectral estimates and geometric parameters
    are descriptive comparisons rather than uncertainty estimates.
    }
    \begin{ruledtabular}
    \begin{tabular}{c c c c c c c c}
        $s$
        & $L_{\rm sp}$
        & $L_{\rm geo}$
        & $\alpha_{\rm sp}$
        & $\alpha_{\rm loc}$
        & $V_b^{\rm sp}$
        & $V_{\rm pk}$
        & $\varepsilon_{\rm ray}$ \\ \hline
        $19$
        & $1.9908$
        & $1.9958$
        & $18.7238$
        & $18.7917$
        & $14.0558$
        & $14.0848$
        & $0.0246$ \\
        $21$
        & $1.9912$
        & $1.9965$
        & $20.7249$
        & $20.7878$
        & $15.5959$
        & $15.5827$
        & $0.0335$
    \end{tabular}
    \end{ruledtabular}
    \label{tab:brane-spectral-reconstruction}
\end{table*}

\begin{figure*}[!htb]
    \centering
    \includegraphics[width=0.95\textwidth]{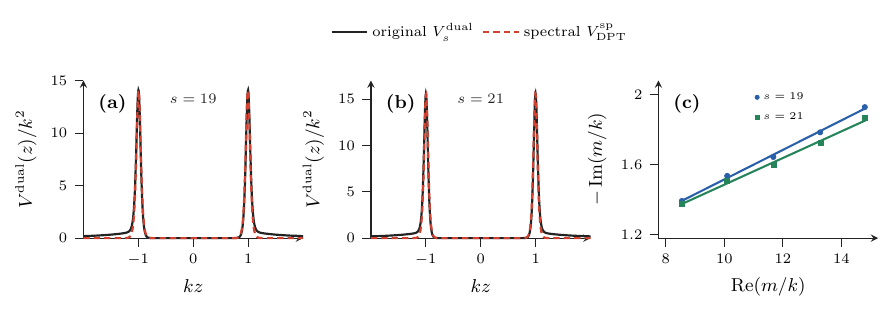}
    \caption{
    Finite-window spectral estimates for split-braneworld dual potentials.
    (a) $s=19$.
    (b) $s=21$.
    Solid curves show the analytic potential in
    Eq.~\eqref{eq:double-brane-dual-potential}.
    Dashed curves show the DPT potential constructed from the
    component height $V_b^{\rm sp}$,
    inverse width $\alpha_{\rm sp}$,
    and center separation $L_{\rm sp}$.
    These parameters are estimated solely from the five Bernstein
    frequencies labeled $n=5,\ldots,9$ here, taken from Table~3
    of Ref.~\cite{Tan:2024dbl},
    without using coordinate-space potential data.
    The parameter estimates, DPT profiles, and spectral rays
    are computed in this work.
    (c) Filled markers show the five input QNMs used for initialization;
    solid lines show the corresponding leading-order sequences
    from Eq.~\eqref{eq:dpt-high-spectrum} with $j_0=0$.
    Their deviations from the markers measure the accuracy
    of the analytic approximation for the input frequencies.
    Frequency and damping predictions for the full DPT potential
    are tested further in Sec.~\ref{subsec:damping}.
    }
    \label{fig:brane-spectral-reconstruction}
\end{figure*}

The real-axis geometric parameters follow from
Eq.~\eqref{barrier-descriptors}.
Denoting the positive-axis peak location by $z_{\rm pk}$,
we have $L_{\rm geo}=2z_{\rm pk}$,
and $\alpha_{\rm loc}$ is calculated from the height
and curvature at that point.
Figure~\ref{fig:brane-spectral-reconstruction}
compares the corresponding potential profiles and spectra.
A small ray residual indicates that the input frequencies
lie close to the analytic scattering sequence;
the predictive accuracy of the full reference potential
requires a further test.

Further spectral fitting to the same five input frequencies,
followed by a direct-integration check for the full DPT potential, gives
\begin{equation}
    (L,\alpha,V_b)
    =
    \begin{cases}
        (1.99873,19.7951,13.5193), & s=19,\\
        (2.00018,22.5285,15.0721), & s=21,
    \end{cases}
    \qquad k=1.
    \label{eq:same-window-refinement}
\end{equation}
The absolute root-mean-square complex-frequency residuals
are $0.02037k$ and $0.02557k$.
Relative to the ray estimates, the inverse widths increase
by $5.7\%$ and $8.7\%$,
the heights change by $-3.8\%$ and $-3.4\%$,
and the separations change by less than $0.5\%$.

At the parameter points given by the ray estimates
and within the frequency windows used,
the cavity condition constructed from exact single-barrier
reflection amplitudes agrees with the QNM frequencies obtained by direct
integration for the full DPT potential to within $4\times10^{-10}k$.
The main corrections from ray estimates to fitted parameters
therefore arise from finite-frequency effects in the reflection amplitude;
PT-component overlap has a negligible effect here.
Even for the same input modes,
$\boldsymbol\theta_{\rm sp}(W)$,
the equal-weight fit $\boldsymbol\theta_{\rm fit}(W,\boldsymbol 1)$,
and $\boldsymbol\theta_{\rm geo}$
represent different parameter descriptions.
The agreement of parameters with real-axis geometry
and the spectral predictive accuracy of the reference potential
must therefore be assessed separately.

\subsection{Higher-Overtone Descriptors and Complex-Pole Benchmarks}
\label{subsec:complex-poles}

The braneworld and DPT potentials have different asymptotic behavior:
$V_s^{\rm dual}\sim3/(4z^2)$,
whereas the DPT potential decays exponentially.
In the pole relations below, lengths are measured in units of $k^{-1}$
and potentials in units of $k^2$. Dimensional frequencies are obtained
by restoring the factor of $k$.
The complex poles of the braneworld potential satisfy $z^{2s}=-1$.
Let $\phi_s=\pi/(2s)$.
Choosing the poles nearest the real axis in the upper-right
and lower-left quadrants gives a displacement $2e^{i\phi_s}$.
Near a pole $z_{p,s}$, the potential has the local expansion
\begin{equation}
    V_s^{\rm dual}(z)
    =
    -\frac{g_s}{(z-z_{p,s})^2}
    +\mathcal O((z-z_{p,s})^{-1}),
    \qquad
    g_s=\frac{3(4s-3)}{16s^2}.
    \label{eq:brane-complex-residue}
\end{equation}

To relate this analytic structure to the DPT reference potential,
we use the complex displacement
$\mathcal L_s=2e^{i\phi_s}$
to define the complex-pole reference parameters
\begin{equation}
    L_{\rm pole}=2\cos\phi_s,
    \qquad
    \alpha_{\rm pole}
    =\frac{\pi}{2\sin\phi_s},
    \qquad
    V_{\rm pole}=g_s\alpha_{\rm pole}^2.
    \label{eq:pole-benchmarks}
\end{equation}
These parameters are calculated directly from the known background
by matching the relevant pole displacement
and the coefficient of the double-pole term.
Thus, $\boldsymbol\theta_{\rm pole}$ characterizes the local
complex analytic structure, while $\boldsymbol\theta_{\rm geo}$
characterizes real-axis peak heights, curvatures, and separation.

If the diagonal phase-balance condition discussed above applies,
this complex displacement gives the candidate spectral spacing
$\Delta m_{\rm pole}=\pi e^{-i\phi_s}/2$.
For $s=19,21$, the corresponding values are
$(1.56543-0.12972i)k$
and $(1.56640-0.11739i)k$, respectively.
Matching $\mathcal L_s$ to $L+i\pi/\alpha$ gives
$(L,\alpha)=(1.9932,19.0217)$
and $(1.9944,21.0196)$.
This matching supplies complex optical scales for comparison
with spectral estimates.
The full resonance condition also involves local connection relations,
more distant poles, and the global analytic continuation to infinity.

Additional direct-integration calculations extend the tracked QNM sequence
to $n=24$.
In the $n=20,\ldots,24$ windows,
the fitted complex spacings differ from the pole-displacement predictions
by less than $8\times10^{-5}k$.
Under the same local-connection assumption,
the double-pole coefficient also predicts the transverse offset.
Using $\lambda=1-a_s$ in the local Bessel/PT connection factor gives
\begin{equation}
    g_s=a_s(1-a_s),
    \qquad
    a_s=\frac{3}{4s},
    \qquad
    \chi_{\rm pole}
    =\ln[4\sin^2(\pi a_s)].
    \label{eq:pole-strength-intercept}
\end{equation}
For $s=19,21$, this yields
$\chi_{\rm pole}=-2.793617,-2.992853$.

Table~\ref{tab:higher-brane-windows}
tests this transverse-offset prediction using the fixed analytic
complex displacement $\mathcal L_s$.
As the selected windows move to higher overtones,
the offset differences decrease to approximately $0.0056$ and $0.0051$.
Over the finite overtone ranges examined,
both the complex spacing and transverse offset therefore support
the interpretation in terms of the nearest complex poles.

Table~\ref{tab:brane-spectral-reconstruction} reports the
Bernstein-based estimates for $n=5$--$9$.
We assess window dependence by comparing the three direct-integration
windows in Table~\ref{tab:higher-brane-windows}.
For $s=19$, as the window moves from $n=10$--$14$ to $n=20$--$24$,
$\alpha_{\rm sp}$ rises from $18.9185k$ to $19.0118k$ and
$V_b^{\rm sp}$ falls from $13.8997k^2$ to $13.7833k^2$.
For both $s=19$ and $s=21$, the absolute fractional deviations of
the inverse width and height from the complex-pole benchmarks decrease
across these windows, and their deviations from the real-axis benchmarks
increase.
A reduction in finite-frequency corrections at higher overtones
therefore does not directly imply improved estimates
of real-axis geometry.
The reference parameters used to define a parameter deviation
must be stated explicitly.
The available windows support this interpretation.
Determining the high-frequency limit of the sequence
and its global Stokes connections requires further analysis
of the algebraic tail and competing connection paths.

\begin{table}[htb]
\caption{
Higher-overtone ray descriptors from direct-integration spectra
compared with complex-pole benchmarks, in units where $k=1$.
The transverse-offset test uses the fixed analytic complex length
$\mathcal L_s=2e^{i\phi_s}$, with
$\chi_W
=2\langle\operatorname{Im}(m_n\mathcal L_s)\rangle_W$
and $\delta\chi=\chi_W-\chi_{\rm pole}$.
The ray parameters $L_{\rm sp}$, $\alpha_{\rm sp}$,
and $V_b^{\rm sp}$ are determined from the complex spacing
and transverse offset fitted in each window.
The quantity $\chi_W$ uses the fixed $\mathcal L_s$
to test the analytic pole benchmark.
}
\begin{ruledtabular}
\begin{tabular}{ccccccc}
$s$ & $n$ window
& $L_{\rm sp}$
& $\alpha_{\rm sp}$
& $V_b^{\rm sp}$
& $\chi_W$
& $\delta\chi$ \\
19 & 10--14
& 1.99318 & 18.9185 & 13.8997
& $-2.778925$ & $0.014692$ \\
19 & 15--19
& 1.99313 & 19.0000 & 13.8079
& $-2.785991$ & $0.007626$ \\
19 & 20--24
& 1.99313 & 19.0118 & 13.7833
& $-2.788043$ & $0.005573$ \\
21 & 10--14
& 1.99442 & 20.8455 & 15.4464
& $-2.975994$ & $0.016859$ \\
21 & 15--19
& 1.99439 & 20.9878 & 15.3165
& $-2.985466$ & $0.007387$ \\
21 & 20--24
& 1.99437 & 21.0076 & 15.2822
& $-2.987709$ & $0.005144$ \\
\end{tabular}
\end{ruledtabular}
\label{tab:higher-brane-windows}
\end{table}

\subsection{Mode-Set Dependence and Predictive Accuracy}
\label{subsec:checks}

Let $\delta m_n=m_n^{\rm DPT}-\widehat m_n$ be the
complex-frequency residual for the matched mode with input index $n$.
For a nonempty index set $\mathcal I$, define the absolute root-mean-square
complex-frequency error
\begin{equation}
    \varepsilon_{\mathcal I}
    =
    \left[
    \frac{1}{|\mathcal I|}
    \sum_{n\in\mathcal I}
    |\delta m_n|^2
    \right]^{1/2}.
    \label{eq:rms-definition}
\end{equation}
This quantity has dimensions of frequency and is reported below
in units of $k$.
Relative complex-frequency errors and relative damping errors
are assessed separately.

For $s=19$, an equal-weight joint fit to the published
Bernstein frequencies with $n=0,1,2,5,6,7$ gives
$(L,\alpha,V_b)=(2.00146,17.8305,13.2799)$,
with an RMS fitting residual $\varepsilon_{\rm fit}=0.0369k$.
For the modes excluded from the fit, $n=3,4,8,9$,
the validation RMS error is $\varepsilon_{\rm val}=0.0424k$.
A separate fit using the published direct-integration frequencies
gives $(L,\alpha,V_b)=(1.99933,18.1861,13.5435)$,
with fitting and validation RMS errors of $0.0329k$ and $0.0222k$,
respectively.
Table~\ref{tab:mode-residuals} lists the real and imaginary residuals
mode by mode to show component differences beyond the aggregate RMS errors.

\begin{table}[htb]
\caption{
Mode-by-mode frequency residuals $\delta m_n$
for the $s=19$ joint fits, in units of $k$.
B and D denote Bernstein and direct-integration inputs, respectively.
The two pairs of columns correspond to separate equal-weight fits.
Modes $n=3,4,8,9$ are excluded from the fits.
}
\begin{ruledtabular}
\begin{tabular}{ccccc}
$n$
& $\operatorname{Re}\delta m_{\rm B}$
& $\operatorname{Im}\delta m_{\rm B}$
& $\operatorname{Re}\delta m_{\rm D}$
& $\operatorname{Im}\delta m_{\rm D}$ \\
0  & $0.00254$  & $-0.05492$ & $0.00308$  & $-0.05551$ \\
1  & $0.03677$  & $-0.00997$ & $0.04242$  & $-0.01987$ \\
2  & $0.04292$  & $-0.02036$ & $0.03015$  & $0.00179$ \\
3  & $0.00652$  & $-0.01139$ & $0.01829$  & $0.00685$ \\
4  & $0.03028$  & $0.02256$  & $0.00891$  & $0.00702$ \\
5  & $-0.00529$ & $0.02086$  & $0.00069$  & $0.00521$ \\
6  & $0.00029$  & $0.01630$  & $-0.00704$ & $0.00236$ \\
7  & $-0.01954$ & $-0.01867$ & $-0.01456$ & $-0.00114$ \\
8  & $-0.06256$ & $-0.01842$ & $-0.02198$ & $-0.00510$ \\
9  & $-0.03375$ & $-0.01381$ & $-0.02930$ & $-0.00936$ \\
\end{tabular}
\end{ruledtabular}
\label{tab:mode-residuals}
\end{table}

Differences between the B and D residuals reflect changes
in both the input frequencies and the fitted parameters.
For example, at $n=1$, the imaginary parts of the two published inputs
differ by $0.00881k$, close to the corresponding difference
of $0.00990k$ between the imaginary residuals.
The rounded tabulated data in Ref.~\cite{Tan:2024dbl}
do not provide per-mode error estimates.
Our integration-contour tests show that the adopted
direct-integration frequencies remain stable at the precision
listed in the Appendix.
Assessing the convergence of the Bernstein frequencies
would require resolution tests for that method.
The approximately $2\%$ changes in fitted $\alpha$ and $V_b$
reflect sensitivity to the method used to compute the input frequencies
and should be distinguished from numerical variations induced
by changing the integration configuration.
Equal-weight fits to both inputs retain a fundamental-mode
damping bias of approximately $35\%$.

To examine variation within the same background family,
we add $s=7,11,15$ and use the same direct-integration implementation
to recompute the tracked modes $n=0,\ldots,9$ for
$s=7,11,15,19,21$.
This comparison uses the recomputed braneworld frequencies throughout,
with the same fitting and validation sets.
The results are given in Table~\ref{tab:additional-s}.

To accelerate optimization, the fitting stage uses the cavity condition
constructed from exact single-barrier reflection amplitudes.
All ten frequencies are then checked using the full summed DPT potential.
For these parameter points and mode ranges,
the frequencies from the two forward calculations differ
by less than $3.2\times10^{-10}k$,
showing that the cavity condition is sufficiently accurate
over the tested range.

\begin{table}[htb]
\caption{
Joint fits for different members of the same braneworld background family.
The table reports numerical values of
$kL$, $\alpha/k$, $V_b/k^2$, and $\varepsilon/k$.
The fitting modes are $n=0,1,2,5,6,7$,
and the validation modes are $n=3,4,8,9$.
The signed relative fundamental-mode damping error is
$\delta\gamma_0/\gamma_0$, where
$\gamma_0=-\operatorname{Im}\widehat m_0$
and
$\delta\gamma_0
=-\operatorname{Im}m_0^{\rm DPT}-\gamma_0$.
The case $s=3$ is omitted from this comparison
because modes with $n>3$ have not been reliably verified,
preventing the use of the same fitting and validation sets.
}
\label{tab:additional-s}
\begin{ruledtabular}
\begin{tabular}{ccccccc}
$s$ & $L$ & $\alpha$ & $V_b$
& $\varepsilon_{\rm fit}$
& $\varepsilon_{\rm val}$
& $\delta\gamma_0/\gamma_0$ \\
7  & 1.95989 & 6.9481  & 4.8709
& 0.0473 & 0.0396 & 38.6\% \\
11 & 1.98791 & 10.7785 & 7.8258
& 0.0373 & 0.0304 & 36.4\% \\
15 & 1.99611 & 14.5341 & 10.7245
& 0.0341 & 0.0251 & 35.8\% \\
19 & 1.99933 & 18.1864 & 13.5436
& 0.0329 & 0.0222 & 35.7\% \\
21 & 2.00021 & 19.9653 & 14.9169
& 0.0326 & 0.0213 & 35.7\% \\
\end{tabular}
\end{ruledtabular}
\end{table}

We also compare three windows containing the same number of modes:
$n=3$--$7$, $4$--$8$, and $5$--$9$.
Across the five backgrounds and three windows,
the maximum absolute relative deviations of the ray parameters
from $(L_{\rm geo},\alpha_{\rm loc},V_{\rm pk})$
are $0.98\%$, $15.95\%$, and $3.50\%$,
in the order $L,\alpha,V_b$.
For each background, we also subtract the smallest window estimate
from the largest and divide by the corresponding real-axis benchmark.
The maximum relative ranges across the five backgrounds
are $0.114\%$, $10.0\%$, and $3.45\%$, respectively.
Thus, within this background family and these windows,
the separation estimate has much weaker window dependence
than the inverse-width estimate.

For the less separated $s=3$ case,
fitting the first three modes with the full DPT spectrum gives
$\varepsilon_{\rm fit}=0.0667k$,
whereas the prediction error for the excluded mode $n=3$ is $0.1571k$.
This example further shows that predictive accuracy must be tested
using frequencies excluded from the fit;
the fitting residual alone does not determine
the range of applicability of the reference potential.

The larger window variation of the inverse width can be understood
from the different responses of the ray map to the complex spacing.
When $\Delta_I/\Delta_R\ll1$,
Eq.~\eqref{eq:inversion-sensitivity} shows that
$\alpha_{\rm sp}$ remains sensitive to relative changes in $\Delta_I$,
while $L_{\rm sp}$ is controlled mainly by changes in $\Delta_R$.
Finite-frequency reflection affects both phase delay and leakage
[Eqs.~\eqref{eq:dpt-large-phase-delay}
and~\eqref{eq:dpt-reflection-subleading}],
so changes in inverse width affect both oscillation frequency and damping.

The least-constrained direction of the joint fit,
discussed in Sec.~\ref{subsec:identifiability},
indicates that variations in inverse width and height
can partially compensate each other.

The following two tests use the consistently recomputed
$s=19,21$ inputs and their initial ray estimates
to examine finite-window bias and input-noise propagation separately.
First, at the corresponding initial parameter points,
we generate synthetic spectra with known parameters using
the cavity model constructed from exact single-barrier reflection
amplitudes and apply the ray inversion map.
Even though the inputs are generated by the same reference
scattering model, the inverse-width estimates from $n=5$--$9$
have biases of $-4.15\%$ and $-5.39\%$,
while the separation biases are $+0.122\%$ and $+0.103\%$.
Moving to $n=40$--$44$ reduces the absolute relative inverse-width
bias below $0.021\%$.

Second, we add independent Gaussian perturbations with standard
deviation $0.005k$ to every real and imaginary component
of the $n=5$--$9$ input frequencies.
Across 10,000 trials, the relative standard deviation
is approximately $0.10\%$ for the separation,
$1.16$--$1.24\%$ for the inverse width,
and $0.53$--$0.59\%$ for the height.
These values describe conditional noise propagation under the assumed
input-error model.
Reference-potential mismatch and deterministic finite-window bias
must be assessed separately.
The synthetic-window biases are listed in the Appendix;
the noise intervals are provided in the accompanying data.

The equal-weight spectral fits above use bounded trust-region
least squares, assigning equal weights to real and imaginary residuals.
Fits to the published inputs use
$1.7\le kL\le2.3$,
$8\le\alpha/k\le35$,
and $6\le V_b/k^2\le25$,
covering a broad parameter range around the ray estimates.
Fits for the additional values of $s$ use
$[1.5,2.6]\times[2,40]\times[1,30]$
in coordinates $(kL,\alpha/k,V_b/k^2)$.

For the $s=19$ Bernstein input,
we also use four different initial guesses and relax the bounds to
$[1.5,2.6]\times[6,45]\times[3,35]$.
The maximum relative range of the resulting parameters
is below $5\times10^{-7}$,
and all optima lie inside the bounds.
The fitted parameters are therefore stable over the tested
initial guesses and parameter bounds.

\subsection{Damping Predictions and the Applicability of the Reference Potential}
\label{subsec:damping}

We now examine the predictive accuracy of the three-parameter
reference potential for different frequency components.
For the $s=19$ Bernstein joint fit,
the four modes excluded from the fit, $n=3,4,8,9$,
have relative complex-frequency errors $|\delta m_n|/|\widehat m_n|$
ranging from $0.24\%$ to $0.53\%$ after rounding.
The RMS of these four relative errors is $0.40\%$.
The reference potential thus predicts these complex frequencies
with good accuracy, but this accuracy does not directly characterize
the relative error in the fundamental-mode damping rate.

For the same fit, the predicted fundamental damping rate is $0.21042k$,
compared with the input value $0.15550k$.
The absolute error is $0.05492k$ and the relative error is $35.3\%$.
The equal-weight fit to the direct-integration input gives
$0.05551k$ and $35.7\%$, respectively.
Since the amplitude decay time is $\tau_0=1/\gamma_0$,
the Bernstein fit predicts $k\tau_0=4.753$,
compared with $6.431$ from the input frequency,
underestimating the lifetime by $26.1\%$.
The actual waveform also depends on mode excitation amplitudes,
mode interference, and the continuum response.
Time-domain tests are therefore needed in addition to frequency fitting.

The approximately $35\%$ damping bias corresponds to the equal-weight
objective above and changes when the fitting weights are varied.
To examine this dependence, consider
\begin{equation}
    \mathcal E_w
    =
    \sum_{n\in\mathcal I_{\rm fit}}
    |\delta m_n/k|^2
    +(w-1)
    [\operatorname{Im}(\delta m_0/k)]^2,
    \qquad w\ge1.
    \label{eq:damping-weight}
\end{equation}
We use the same six fitting modes and four validation modes.
Only the weight of the fundamental-mode imaginary part is changed;
all other frequency components retain their original weights.
To compare results at different $w$,
Table~\ref{tab:weights} reports unweighted fitting
and validation RMS errors throughout.

Increasing $w$ from 1 to 1024 reduces the relative fundamental
damping error for the Bernstein input from approximately $35\%$
to $0.66\%$.
However, the validation RMS error increases from $0.0424k$ to $0.1253k$,
approximately three times its original value.
The relative RMS error of the four validation modes also increases
from $0.40\%$ to $1.36\%$,
the relative error in the real part of the fundamental frequency
grows to approximately $9\%$,
and the inverse-width parameter decreases by approximately $24\%$.

The recomputed inputs exhibit the same accuracy trade-off:
the relative damping error decreases from $35.71\%$ to $0.67\%$,
while the validation RMS error increases from $0.0222k$ to $0.1162k$.
Representative weights are listed in the table;
all intermediate weights and mode-by-mode residuals
are included in the accompanying data.
For $w=1024$, a different initial guess gives relative parameter
differences below $1.4\times10^{-7}$,
and the optimum lies inside the relaxed bounds.

These results show that changing the relative fitting weights
alters the balance between the spectral features reproduced
by the three-parameter reference potential.
Under the illustrative criteria in
Sec.~\ref{subsec:inversion-validation},
none of the six tested weights simultaneously reduces
the validation complex-frequency RMS error below $0.05k$
and the relative fundamental damping error below $10\%$.
This result describes the accuracy trade-off for the current
input data, mode sets, and fitting setup.

\begin{table}[htb]
\caption{
Damping-weight tests for the published $s=19$ Bernstein input.
The table reports numerical values of
$kL$, $\alpha/k$, $V_b/k^2$, and $\varepsilon/k$;
the signed relative damping error is expressed as a percentage.
Both RMS-error columns are unweighted
and use the original fitting and validation sets.
At these optimal parameter points, frequencies from the full DPT
potential and the cavity condition differ
by less than $3.1\times10^{-11}k$.
}
\label{tab:weights}
\begin{ruledtabular}
\begin{tabular}{ccccccc}
$w$ & $L$ & $\alpha$ & $V_b$
& $100\delta\gamma_0/\gamma_0$
& $\varepsilon_{\rm fit}$
& $\varepsilon_{\rm val}$ \\
1
& 2.00146 & 17.8305 & 13.2799
& 35.32 & 0.0369 & 0.0424 \\
16
& 2.00038 & 15.5119 & 13.1206
& 19.12 & 0.0565 & 0.0746 \\
64
& 1.99979 & 14.2513 & 13.2058
& 8.09 & 0.0819 & 0.1039 \\
1024
& 1.99948 & 13.5017 & 13.3394
& 0.66 & 0.1013 & 0.1253 \\
\end{tabular}
\end{ruledtabular}
\end{table}

We next use deformations of auxiliary scattering potentials
to examine the sensitivity of damping to the outer potential profile.
We compare exterior replacements that change the asymptotic tail
with compactly supported deformations that preserve it.
These potentials diagnose spectral sensitivity
and need not correspond to self-consistent braneworld background solutions.

In the deformation tests we take $k=1$, so lengths are given in
units of $k^{-1}$. The reference DPT profile uses
$(kL,\alpha/k,V_b/k^2)=(2.00146,17.8303,13.2798)$.
Replacing the exterior by this profile with a smooth transition over
$1.15\le |z|\le 1.5$ increases the fundamental damping rate by $19.85\%$.
The central-profile replacement over the transition interval
$0.7\le |z|\le 0.9$ changes the damping by $-0.030\%$.
The exterior replacement changes both the potential shoulders
and the asymptotic decay.
Hankel and plane-wave outgoing conditions are used
as appropriate for the respective tails.
This comparison therefore probes the overall exterior profile,
including changes in both the outer potential shape
and its asymptotic behavior.

To examine profile effects while preserving the original asymptotic tail,
we introduce a compactly supported deformation with dimensionless
amplitude $\xi$
\begin{equation}
    V_\xi
    =
    V_{19}^{\rm dual}
    +\xi\,\mathcal S_{1.15,1.5}(|z|)
    [1-\mathcal S_{R_{\rm restore},R_{\rm restore}+0.3}(|z|)]
    (V_{\rm DPT}-V_{19}^{\rm dual}),
    \qquad 0\le \xi\le1.
    \label{eq:controlled-profile}
\end{equation}
The switch $\mathcal S_{a,b}(|z|)$ vanishes for $|z|\le a$,
equals one for $|z|\ge b$, and takes the form
$10\zeta^3-15\zeta^4+6\zeta^5$ between the endpoints, where
$\zeta=(|z|-a)/(b-a)$.
The radius $R_{\rm restore}$ marks the start of the transition
back to the original braneworld potential. For
$|z|\ge R_{\rm restore}+0.3$, the potential coincides with the original.
All cases therefore retain the algebraic asymptotic tail $3/(4z^2)$
and use the same Hankel outgoing condition.

Starting from the fundamental mode at $\xi=0$,
we follow its QNM branch continuously as $\xi$ varies.
Table~\ref{tab:local-profile} reports the resulting relative changes
in the damping rate for three values of $R_{\rm restore}$.
Changing the integration contour and tightening the integration tolerances
changes the frequencies by less than $9\times10^{-13}k$.

\begin{table}[htb]
\caption{
Relative changes
$[\gamma_0(\xi,R_{\rm restore})-\gamma_0(0)]/\gamma_0(0)$
in the damping rate along the QNM branch continued from
the fundamental mode at $\xi=0$,
induced by the compactly supported deformation
in Eq.~\eqref{eq:controlled-profile}.
The results are evaluated at $k=1$ and expressed as signed percentages.
All cases preserve the original algebraic asymptotic tail,
and all changes vanish at $\xi=0$.
}
\label{tab:local-profile}
\begin{ruledtabular}
\begin{tabular}{cccc}
$\xi$ & $R_{\rm restore}=2$ & $R_{\rm restore}=3$ & $R_{\rm restore}=5$ \\
0.25 & $+7.61$  & $+6.11$  & $+6.55$ \\
0.5  & $+15.95$ & $+11.10$ & $+13.58$ \\
0.75 & $+25.00$ & $+14.74$ & $+21.11$ \\
1    & $+34.63$ & $+17.05$ & $+28.93$ \\
\end{tabular}
\end{ruledtabular}
\end{table}

The damping rate along this branch can change substantially
even when the asymptotic structure is unchanged.
The approximately $17$--$35\%$ changes at $\xi=1$,
together with their nonmonotonic dependence on $R_{\rm restore}$,
show that the outer potential profile
has an important effect on resonance leakage.
Thus, even with the peak geometry and asymptotic tail fixed,
the profile outside the peaks contains scattering information
that is important for damping.
These deformations test sensitivity to the potential shape;
their damping shifts should not be interpreted as additive components
of the full DPT fitting error.

The effect of damping bias on a single-mode envelope
can be estimated directly.
For the $s=19$ equal-weight fit to the Bernstein frequencies, let
$\gamma_{\rm fit}=-\operatorname{Im}m_0^{\rm DPT}$ and
$\gamma_{\rm ref}=-\operatorname{Im}\widehat m_0$.
For isolated modes with equal initial amplitudes, the ratio of the
fitted envelope to the reference envelope at $t=1/\gamma_{\rm ref}$ is
$e^{-(\gamma_{\rm fit}-\gamma_{\rm ref})/\gamma_{\rm ref}}
\simeq0.70$.
The time for which the envelope remains above the same amplitude
threshold is approximately $26\%$ shorter,
potentially changing the late-time contribution
to the accumulated signal-to-noise ratio.
Echo spacing is determined by propagation times
and reflection delays and requires the corresponding
phase or time-domain analysis.

Table~\ref{tab:weights} illustrates fitting-accuracy trade-offs
between different spectral features.
Choosing fitting weights to optimize a detector-noise-weighted
waveform match would additionally require the mode excitation
amplitudes and phases, together with the detector noise
power spectral density.

The model with a finite extra dimension in Ref.~\cite{Jia:2024finite}
provides an example of a different trapping mechanism.
Its warp function is $A(y)=\nu\ln[\cos(ky)]$, where $y$ is the
proper-distance coordinate along the extra dimension and $\nu>0$ is
a dimensionless parameter.
In the tensor sector, the continuum starts at $m^2=0$
for $\nu>1$.
For $\nu=1$, the continuum threshold is $9k^2/4$,
and a massive bound state exists at $m^2=2k^2$.
This model has a different spectral threshold and bound-state structure,
so its trapping mechanism must be analyzed within the corresponding
background.
Inferring brane splitting from long-lived gravitational excitations
therefore requires examining the potential shape, spectral threshold,
and trapping mechanism together to establish the physical meaning
of the extracted distance scale.

\section{Conclusion}\label{conclusion}

Using an established analytic thick-brane family and the symmetric
double-PT reference potential, we developed a procedure to assess
the accuracy of effective-potential inference from finite QNM spectra
and quantify the associated systematic errors.
We derived the closed-form quadratic shift for overlapping barriers
at fixed overtone number and used the independent-barrier
high-frequency relation to initialize the inversion. Direct
integration with outgoing boundary conditions tested the predictions
of these approximations for the full summed potential. Synthetic
spectra, local sensitivity tests, and modes excluded from the fit
provide checks of finite-window bias, parameter identifiability,
and spectral predictive accuracy.

For the high-frequency ray approximation, the relevant scale checks
are $|m|/\alpha_{\rm PT}\gg1$ and $|m|L\gg1$ throughout the selected
window [Eq.~\eqref{eq:dpt-resolution-condition}], within the sector
specified for Eq.~\eqref{eq:dpt-reflection-subleading}. In the synthetic
cavity-model tests, the absolute relative inverse-width bias
decreases from $4.15\%$--$5.39\%$ for $n=5$--$9$ to below $0.021\%$
for $n=40$--$44$ (see Sec.~\ref{subsec:synthetic}). The conditioning
of the inverse problem also depends on the window. At the representative
fitted point, in the adopted logarithmic parameter coordinates,
shifting from $n=5$--$9$ to $n=20$--$24$ raises the condition number
from $68.3$ to $739.9$ and reduces the smallest singular value.
In the upper window, the least-constrained parameter combination is
dominated by the barrier height (Table~\ref{tab:jacobian}).

For each inferred parameter, we track its absolute fractional
deviation from each of the two benchmarks as we shift equal-size
overtone windows. A decreasing deviation from the complex-pole value
and an increasing deviation from the real-axis geometric value
identify a shift toward the pole scale across those windows. For $s=19,21$, the ray
estimates of inverse width and height show this behavior in the
direct-integration spectra as the window moves from $n=10$--$14$
to $n=20$--$24$
(Table~\ref{tab:higher-brane-windows}). The separation estimate varies
less across the tested windows. In the highest window, the complex
spacings agree with the pole-based predictions to within
$8\times10^{-5}k$, and the transverse-offset deviations decrease
across the tested windows. Establishing the asymptotic braneworld
spectrum requires a global analysis of the Stokes connections.

The choice of fitting weights changes which spectral quantities the
reference potential predicts most accurately. For the $s=19$
Bernstein input, an equal-weight fit gives a relative RMS
complex-frequency error of $0.40\%$ on four modes excluded from the
fit and a relative error of $35.3\%$ in the fundamental-mode damping
rate. Increasing the damping weight to $1024$ reduces the damping
error to $0.66\%$ and raises the relative validation RMS error to
$1.36\%$. The error in the fundamental oscillation frequency also
increases. Compactly supported deformations outside the peak region
produce substantial damping shifts with the algebraic tail fixed,
showing the sensitivity of resonance leakage to the outer profile.

Finite-spectrum constraints should be reported with the reference
family, overtone window, mode assignment, and fitting weights, along
with the benchmark or observable used to assess accuracy. Decay
times and waveforms require validation with error measures appropriate
to those observables. Connecting effective-potential constraints to
microscopic brane structure requires a specified background model
and additional physical information.

\appendix
\section{Numerical Methods and Accuracy Tests}
\label{subsec:numerical-method}

\subsection{Direct Integration and Numerical Accuracy Checks}
\label{subsec:outgoing-validation}

We use direct integration to compute the selected QNM sequences,
imposing outgoing boundary conditions at the asymptotic ends.
The tests below check the identification and numerical stability
of the tracked roots over the frequency and parameter ranges examined.

For the DPT potential, we write the outgoing solution on the positive
half-axis as $\psi=e^{imz}f$, which gives
\begin{equation}
    f''+2imf'-V_Lf=0.
    \label{eq:factored-ode}
\end{equation}
At a finite endpoint on the complex contour, we impose the leading
asymptotic data $f=1$ and $f'=0$ and integrate toward the center
along an exterior complex ray followed by a real-axis segment.
Even- and odd-parity QNMs satisfy
$f'(0)+imf(0)=0$ and $f(0)=0$, respectively.
We integrate the factored equation with an eighth-order explicit
Runge--Kutta method (DOP853).

For the braneworld potential, boundary data on the positive half-axis
are obtained from $\sqrt zH_1^{(1)}(mz)$,
the outgoing solution of the asymptotic equation with
$V=3/(4z^2)$.
For the compactly supported deformation in
Eq.~\eqref{eq:controlled-profile},
the exterior complex ray begins outside the support of the deformation,
and the deformed region is traversed along the real axis.

We test the numerical stability of the frequencies by varying
the starting point, length, and rotation angle of the exterior
complex ray, together with the integration tolerances.
The main settings are listed in
Table~\ref{tab:integration-settings}.
Let $\varphi=\arg(m^*)$ and define
\[
    \Theta_d
    =\min\!\left(1.45,\max(0.2,\varphi+d)\right),
    \qquad
    \Xi_d
    =\min\!\left(1.35,\max(0.2,\varphi+d)\right).
\]
Absolute tolerances, settings for the two-sided integrations,
and implementation details are provided in the accompanying scripts.

\begin{table*}[htb]
\caption{
Integration settings used in the numerical stability tests.
Paired entries give the baseline and comparison settings,
respectively; a single entry is shared by both configurations.
DPT lengths are measured in units of $\alpha_{\rm PT}^{-1}$,
and braneworld lengths in units of $k^{-1}$.
$R_c$ and $T$ are the starting point and length of the exterior
complex ray, respectively.
Rotation angles are in radians, and the listed tolerances are relative.
The restoration radius $R_{\rm restore}$ in
Eq.~\eqref{eq:controlled-profile} is expressed here in the
dimensionless integration coordinate.
}
\label{tab:integration-settings}
\begin{ruledtabular}
\begin{tabular}{lcccc}
Calculation & $R_c$ & $T$ & Rotation angle & Relative tolerance \\
DPT high overtones, oscillatory branch
& $\ell/2+\{0.6,0.4\}$
& $80/110$
& $\Theta_{0.12}/\Theta_{0.18}$
& $10^{-9}/10^{-11}$ \\
DPT high overtones, shallow-barrier branch
& $\ell/2+\{0.5,0.7\}$
& $70/90$
& $\Theta_{0.23}/\Theta_{0.35}$
& $2\!\times\!10^{-10}/2\!\times\!10^{-11}$ \\
Braneworld $n=0$--$9$
& $2/1.5$
& $80/100$
& $\Xi_{0.35}/\Xi_{0.45}$
& $2\!\times\!10^{-10}/2\!\times\!10^{-11}$ \\
Braneworld $n=10$--$24$
& $2/1.5$
& $80/100$
& $\Xi_{0.35}/\Xi_{0.45}$
& $2\!\times\!10^{-10}/10^{-11}$ \\
Compactly supported deformation
& $R_{\rm restore}+0.8$
& $80/100$
& $0.7/0.8$
& $10^{-11}/2\!\times\!10^{-12}$ \\
\end{tabular}
\end{ruledtabular}
\end{table*}

Under these changes, the high-overtone frequencies on the oscillatory
DPT branch vary by less than
$3.3\times10^{-6}\alpha_{\rm PT}$.
For the shallow-barrier parameter sets used in the applications,
the variations are below $5.2\times10^{-7}k$.
The braneworld calculations reproduce the selected published
direct-integration frequencies with $n=0,1,2,5,7,9$
to within $10^{-5}k$.
For the recomputed sequences $n=0,\ldots,9$
and $n=10,\ldots,24$,
the largest frequency differences between paired configurations
are below $8.2\times10^{-10}k$ and $1.1\times10^{-9}k$,
respectively.
For the compactly supported deformations tested here,
the maximum frequency variation is below $9\times10^{-13}k$.

To cross-check the results obtained from the parity conditions,
we also integrate the unfactored equation
$\psi''=(V-m^2)\psi$ from both ends with the Dormand--Prince
Runge--Kutta method of order $5(4)$ (RK45) and locate zeros of the
normalized Wronskian
at a real-axis matching point:
\begin{equation}
    \widehat W
    =
    \frac{
    \psi_-\psi_+'-\psi_-'\psi_+}
    {
    \|(\psi_-,\psi_-')\|_2
    \|(\psi_+,\psi_+')\|_2}.
\end{equation}
Here $\psi_-$ and $\psi_+$ satisfy the left and right outgoing
conditions, respectively.
In this residual, derivatives are taken with respect to
$\alpha_{\rm PT}z$ for DPT and $kz$ for the braneworld potential.
No central parity condition is imposed.
The normalization leaves the matching zeros unchanged
and makes $|\widehat W|$ independent of the overall amplitudes
of the two solutions.

In the low-overtone tests, shifting the dimensionless real-axis
matching point from $0.17$ to $0.23$,
while also changing the exterior integration contour,
changes the dimensionless frequencies by less than
$1.3\times10^{-13}$.
Table~\ref{tab:independent-check} lists representative differences
between the two-sided matching roots and the roots obtained
from parity conditions, together with the normalized Wronskian
residuals evaluated at the latter.
For the high-overtone tests with $s=19,21$ and $n=20,24$,
the wave function is rescaled between integration segments
to control its magnitude.
The two methods agree to within $9.2\times10^{-11}k$.
Full results for the tested modes are provided in the accompanying data.

\begin{table}[htb]
\caption{
Two-sided matching tests for representative QNMs.
For DPT, $V_b=\alpha_{\rm PT}^{2}$ and $\alpha_{\rm PT}=1$;
for the braneworld background, $k=\delta=1$.
Frequency differences are given in units of $\alpha_{\rm PT}$
and $k$, respectively.
$m_{\rm match}$ denotes the two-sided matching root, and $m_{\rm par}$ denotes
the root obtained from the parity condition.
}
\label{tab:independent-check}
\begin{ruledtabular}
\begin{tabular}{lcc}
Case
& $|m_{\rm match}-m_{\rm par}|$
& $|\widehat W(m_{\rm par})|$ \\
DPT $\ell=0.5,n=1$
& $2.5\times10^{-12}$
& $3.6\times10^{-12}$ \\
DPT $\ell=1.5,n=1$
& $5.0\times10^{-12}$
& $1.1\times10^{-11}$ \\
DPT $\ell=4.0,n=2$
& $3.3\times10^{-12}$
& $2.6\times10^{-11}$ \\
Braneworld $s=19,n=0$
& $1.7\times10^{-12}$
& $4.8\times10^{-12}$ \\
\end{tabular}
\end{ruledtabular}
\end{table}

Roots are tracked at fixed parity as the model parameters vary.
In the braneworld sequence computed using the dual potential,
even $n$ corresponds to even spatial parity and odd $n$ to odd spatial
parity. The superpotential is odd for these backgrounds, so the
intertwining map in Eq.~\eqref{susy-intertwining} reverses the parity
of the corresponding nonzero modes of the partner potentials.
Starting at $s=19$, we obtain the roots for smaller odd values of $s$
by successive continuation through intermediate odd values.
For $s=21$, the published frequencies are used as initial guesses.
Only roots with the parity of the target branch are retained.
For $s=3$, modes $n=0,1,2$ are fitted,
and the verified $n=3$ mode is used for validation.

Spectral fitting uses bounded trust-region least squares.
The parameter bounds for equal-weight fits are given in the main text.
Damping-weighted fits use
$[1.7,2.3]\times[4,80]\times[2,60]$
in coordinates $(kL,\alpha/k,V_b/k^2)$,
and all resulting optima lie inside these bounds.
Finite-difference steps, stopping tolerances,
and initial guesses are specified in the accompanying scripts.
These settings control the numerical integration and optimization;
the predictive error of the reference potential is assessed
through the spectral comparisons in the main text.

\subsection{Finite-Window Bias and Input-Noise Propagation}
\label{subsec:synthetic}

The tests in this subsection use the recomputed
$s=19,21$ braneworld frequencies
and their ray estimates from the $n=5$--$9$ windows.
These inputs differ from the published Bernstein frequencies used in
Table~\ref{tab:brane-spectral-reconstruction}.

For the synthetic-bias test, the two ray estimates are taken
as known parameter values.
We generate noiseless synthetic spectra using the cavity model
constructed from exact single-barrier reflection amplitudes
and then apply the leading-order ray inversion map.
The agreement between this cavity model and the full summed DPT
potential has been checked in the corresponding fitting windows
and for representative higher overtones.
Table~\ref{tab:synthetic-bias} gives the resulting finite-window
parameter biases.
The test quantifies the bias of the leading-order inversion
within the reference scattering model;
it does not include mismatch between the braneworld
and reference potentials.

\begin{table}[htb]
\caption{
Signed synthetic parameter biases
$100(\theta_{\rm ray}/\theta_{\rm true}-1)$,
evaluated componentwise and expressed as percentages.
$\theta_{\rm true}$ denotes the known parameter values used
to generate the noiseless synthetic spectra with the cavity model.
The label $s$ identifies the braneworld spectrum
from which the generating parameter values were estimated.
Units are chosen such that $k=1$.
}
\label{tab:synthetic-bias}
\begin{ruledtabular}
\begin{tabular}{ccccc}
Parameter set & $n$ window
& $L$ bias & $\alpha$ bias & $V_b$ bias \\
19 & 5--9
& $+0.1215$ & $-4.1483$ & $+2.3879$ \\
19 & 20--24
& $+0.0311$ & $-0.0769$ & $+0.3041$ \\
19 & 40--44
& $+0.0087$ & $-0.0195$ & $+0.1795$ \\
21 & 5--9
& $+0.1029$ & $-5.3934$ & $+2.3343$ \\
21 & 20--24
& $+0.0317$ & $-0.0814$ & $+0.2863$ \\
21 & 40--44
& $+0.0089$ & $-0.0202$ & $+0.1686$ \\
\end{tabular}
\end{ruledtabular}
\end{table}

For the input-noise test, independent Gaussian perturbations
with standard deviation $\sigma_{\rm noise}$ are added to each real
and imaginary component of the recomputed
$n=5,\ldots,9$ frequencies.
For each background, we perform 10,000 trials at each of
$\sigma_{\rm noise}=0.001k$ and $0.005k$,
reapplying the ray inversion map in every trial.
The empirical $2.5$th and $97.5$th percentiles
define the central $95\%$ Monte Carlo intervals.
The unperturbed estimates, simulated intervals,
and random-number settings are provided
in the accompanying data and scripts.

These intervals quantify noise propagation conditional on
the selected frequency window, mode assignments,
and Gaussian input-error model.
They do not include finite-window bias,
numerical errors in direct integration,
or reference-potential mismatch,
and are not interpreted as observational confidence intervals.

The main text also compares three windows containing the same
number of modes: $n=3$--$7$, $4$--$8$, and $5$--$9$.
For each background and parameter, we record the minimum
and maximum values of
$100[\theta_{\rm ray}(W)/\theta_{\rm geo}-1]$
across these windows.
These ranges quantify deterministic window dependence.
The corresponding results for all tested backgrounds
are provided in the accompanying data.

\section*{Acknowledgments}

AI-assisted tools were used to refine the language and formatting of the LaTeX manuscript. Responsibility for the scientific content and its validation remains with the authors.

We would like to thank Wen-Yi Zhou for helpful discussions. 
This work was supported by 
the National Natural Science Foundation of China (Grant Nos. 12475056, 12205129, and 12247101),
the Fundamental Research Funds for the Central Universities (Grant Nos. lzujbky-2025-it05 and lzujbky-2025-jdzx07), 
the Natural Science Foundation of Gansu Province (Grant Nos. 22JR5RA389 and 25JRRA799), 
and the ``111 Center'' under Grant No. B20063.

\end{document}